\documentclass[twocolumn]{aastex631}

\newcommand{\unit}[2]{\ensuremath{\textrm{#1}^{#2}}}

\usepackage{hyperref}

\shorttitle{Orientations of DM Halos in Latte Galaxies}
\shortauthors{Baptista et al.}

\begin{document}

\correspondingauthor{Jay Baptista}
\email{jay.baptista@yale.edu}

\title{Orientations of DM Halos in FIRE-2 Milky Way-mass Galaxies}


\newcommand{\Yale}{\affiliation{Department of Astronomy and Astrophysics, Yale University, New Haven, CT 06520, USA}}

\newcommand{\IfA}{\affiliation{Institute for Astronomy, University of Hawai`i, 2680 Woodlawn Drive, Honolulu, HI 96822, USA}}

\newcommand{\Penn}{\affiliation{Department of Physics \& Astronomy, University of Pennsylvania, Philadelphia, PA 19104, USA}}

\newcommand{\CCA}{\affiliation{Center for Computational Astrophysics, Flatiron Institute, New York, NY 10010, USA}}

\newcommand{\NU}{\affiliation{Department of Physics and Astronomy and CIERA, Northwestern University, 1800 Sherman Ave, Evanston, IL 60201, USA}}

\newcommand{\UT}{\affiliation{Department of Astronomy, The University of Texas Austin, 2515 Speedway, Stop C1400, Austin, TX 78712, USA}}

\newcommand{\UA}{\affiliation{Department of Physics and Astronomy, University of Alabama, Box 870324, Tuscaloosa, AL 35487, USA}}

\newcommand{\Caltech}{\affiliation{TAPIR, California Institute of Technology, MC 350-17, Pasadena, CA 91125, USA}}
\author[0000-0002-9306-1704]{Jay Baptista}
\Yale
\IfA

\author[0000-0003-3939-3297]{Robyn Sanderson}
\Penn
\CCA

\author[0000-0001-8832-4488]{Dan Huber}
\IfA

\author[0000-0003-0603-8942]{Andrew Wetzel}
\affiliation{Department of Physics and Astronomy, University of California, Davis, CA 95616, USA}

\author[0000-0003-4394-6085]{Omid Sameie}
\UT

\author[0000-0002-9604-343X]{Michael Boylan-Kolchin}
\UT

\author[0000-0001-6380-010X]{Jeremy Bailin}
\UA

\author[0000-0003-3729-1684]{Philip F. Hopkins}
\Caltech

\author[0000-0002-4900-6628]{Claude-André Faucher-Giguere}
\NU

\author[0000-0001-6711-8140]{Sukanya Chakrabarti}
\affiliation{School of Physics and Astronomy, University of Alabama, Huntsville, 301 Sparkman Drive, Huntsville, AL 35899, USA}
\affiliation{Institute of Advanced Study, 1 Einstein Drive Princeton, New Jersey 08540, USA}

\author{Drona Vargya}
\Penn

\author[0000-0001-5214-8822]{Nondh Panithanpaisal}
\Penn

\author[0000-0002-8354-7356]{Arpit Arora}
\Penn

\author[0000-0002-6993-0826]{Emily Cunningham}
\CCA

\begin{abstract}

The shape and orientation of dark matter (DM) halos are sensitive to the micro-physics of the DM particle, yet in many mass models, the symmetry axes of the Milky Way's DM halo are often assumed to be aligned with the symmetry axes of the stellar disk. This is well-motivated for the inner DM halo (within a few disk scale radii) but not for the outer halo.
We use zoomed cosmological-baryonic simulations from the Latte suite of FIRE-2 Milky Way-mass galaxies to explore the evolution of the DM halo's orientation with radius and time, in the presence or absence of a major merger with a Large Magellanic Cloud (LMC) analog, and when varying the DM model. 
In three of the four cold DM halos we examine, the orientation of the halo minor axis diverges from that of the stellar disk by more than 20 degrees beyond about 30 galactocentric kpc, reaching a maximum of 30--90 degrees depending on the individual halo's formation history. In simulations with identical initial conditions and baryonic physics but using a model of self-interacting DM with $\sigma = 1\ \unit{cm}{2} \unit{g}{-1}$, the halo remains aligned with the stellar disk out to $\sim$200--400 kpc. Interactions with massive satellites ($M \gtrsim 4 \times 10^{10} \, \rm{M_\odot}$ at pericenter; $M \gtrsim 3.3 \times 10^{10} \, \rm{M_\odot}$ at infall) affect the orientation of the halo significantly, aligning the DM halo's major axis with the satellite galaxy from the stellar disk to the virial radius. 
The relative orientation of the halo and disk beyond 30 kpc is thus a potential diagnostic of DM self-interaction if the effects of massive satellites can be accounted for.

\end{abstract}




\section{Introduction}
\label{sec:intro}

Dark matter halos have the potential to serve as macroscopic laboratories that can constrain the micro-physics of a dark matter particle \citep[e.g.][and references therein]{bbk17}. Like all galaxies, the Milky Way (MW) is embedded within a dark matter halo, but despite our close-up view, its structure is not well constrained outside the inner 20 kpc. Especially difficult to constrain are the approximate principal axis ratios of the halo and their orientation relative to the Galactic disk \citep{vera2011, hattori20}. In external galaxies, the geometry of the DM halo is often determined by analyzing weak lensing, polar rings, and the distribution of satellite galaxies. Within our galaxy, methods for constraining the geometry of the dark matter halo rely on modeling the dynamics of distant tracers such as standard-candle stars, tidal streams, globular clusters, or dwarf galaxies, and are thus limited by the depth of current surveys \citep{vera2011, whitepaper, hattori20, gaia, reino21}. 

With the advent of new dynamical information about these tracers, from missions such as Gaia, JWST, Roman, and Rubin, we will be able to detect stellar tracers out to the edge of the Milky Way's dark matter halo \citep{sanderson+17,whitepaper} and resolve the stars in the faint outskirts of up to 100 MW-like galaxies nearby \citep{pearson+22}. Now is thus the time to examine the geometry of dark matter halo outskirts in realistic simulations of Milky Way analogs in order to make predictions for the shape and orientation of the halo beyond 20 kpc \citep[e.g.][]{prada2019} as well as investigate its sensitivity to the physics of the DM particle \citep[e.g.][]{vargya2021}.

Theoretical studies of the expected dark matter halo shape have been done using simulations of MW-like galaxies both with and without the incorporation of baryonic physics \citep{vera2011, prada2019} and by varying the DM model \citep{vargya2021}. However, somewhat less attention has been paid to the orientation of the symmetry axes of the DM halo as a function of radius and time, especially in cosmological-baryonic simulations where central galaxy formation has been shown to affect DM halo shape \citep{bailin04, allgood2006, gk18, prada2019}. Orientations of the DM halo axes may play an important role in the kinematic twisting of the stellar components in the outer stellar halo of observed galaxies since the potential due to DM mainly drives stellar kinematics at those radii \citep{kormendy09, foster16, pulsoni18, ene18, huang18}.

Furthermore, the ongoing interaction of the Milky Way with the Large Magellanic Cloud (LMC) is expected to influence the orientation of the DM symmetry axes, especially at intermediate radii \citep{gc20, vas21}, yet studies of the geometry of halos in cosmological MW analogs to date do not include LMC-like interactions.

Previous studies using DM-only simulations \citep[e.g.][]{allgood2006,vera2011} indicate that the symmetry axes of halos can twist substantially from the interior to the virial radius, However, models of the Milky Way potential nearly always incorporate the assumption that the dark matter symmetry axes are aligned to the central galaxy at all radii, which can end up producing inconsistent results \citep{law09, debattista, prada2019, vas21}. Although the halo and stellar disk axes are likely to be well-aligned where stars and gas dominate the mass distribution, which includes the inner 20 kpc or so where the MW shape is currently well constrained, there is no reason to assume the halo axes should align to the stellar disk at larger distances or out to the virial radius. \citep{bailin05}  find evidence of decoupling between the disk axes and the halo axes in hydrodynamical simulations beyond $\sim 0.1 R_{\rm vir}$. Analyses of the Milky Way's potential that have been modeled using the assumption of an aligned disk and halo \citep{cunningham} may thus be flawed if the outer halo is tilted significantly relative to the disk. Current observations of the Sagittarius Stream (a powerful tracer of the MW potential) suggest the MW disk is oblique to the outer dark matter halo \citep{debattista,hattori20,vas21}. In the Auriga simulations, simulated halos can become oblique to the symmetry axes of the central galaxy as a function of host distance and time \citep{prada2019}. However, it is reasonable to expect that the radial extent of the alignment between galaxies and their host halos would be somewhat dependent on the nature and strength of stellar and AGN feedback in the galaxy, which can couple small-scale regions in the central galaxy to much larger scales.  It is therefore useful to study the obliquity of DM halos in simulations with a variety of feedback prescriptions, in the interest of isolating these effects. 

In this paper, we use the Latte suite of 13 simulated galaxies ($M^* \sim 5 \times 10^{10} \, \rm{M_\odot}$,  $M_\mathrm{halo} \sim 10^{12} \, \rm{M_\odot}$) \citep{wetzel2016}, which uses substantially different feedback physics than the simulations mentioned above \footnote{See the FIRE project website at \url{http://fire.northwestern.edu}} \citep{fire2}. We also investigate how an in-falling LMC-mass galaxy can dynamically perturb the orientation of the DM halo by analyzing halo orientations throughout several LMC-like mergers with varying masses. Finally, we examine the differences in the relative orientation of the halo and disk when the same halo is simulated with elastic-scattering self-interacting dark matter (SIDM) from the same initial conditions. The SIDM model implemented is modeled as elastic hard-sphere scattering with a fixed cross-section per unit mass, as compared to alternative models that include dissipative SIDM or excited-state SIDM. We perform this analysis to determine whether next-generation surveys could distinguish between CDM and SIDM by constraining the orientation of the DM halo.

This paper is organized as follows. In Section \ref{s.methods}, we discuss the suite of Milky Way-mass simulations we analyze in this paper and the evolutionary histories of each simulation. In Section \ref{ss.symmaxes}, we present a method used to define the symmetry axes of the dark matter halo and the stellar disk. In Section \ref{ss.obliq}, we introduce a procedure to quantify the orientation between the disk and halo based on the halo and disk symmetry axes. In Section \ref{ss.SIDM_DMO}, we introduce our SIDM and dark-matter-only (DMO) halos and outline procedural differences for calculating orientations in those simulations. Section \ref{ss.triax} outlines the process of calculating the triaxiality (shape) of the halo over time as a supplementary measurement to quantify merger effects. In Section \ref{s.ResultsAnddiscussion}, we introduce the halo-disk orientations at the present-day (Section \ref{ss.obliq_z0}) and as a function of redshift (Section \ref{ss.obliq_across_z}). Additionally, we explore the effects of LMC-mass satellites on orientation (Section \ref{ss.obliq_across_lmc}). In Section \ref{ss.sidm_dmo_results}, we perform our analysis of SIDM and DMO halo-disk orientations at the present day and present the differences in results between the simulations. In Section \ref{ss.short_axis_filaments}, we discuss the orientation of the halo short axis relative to the enclosing dark matter filament across our different simulations. We conclude our results and discussion in Section \ref{s.conclusions}.

\setcounter{table}{0}
\begin{deluxetable*}{lccccccccccccr}
\label{tab.sims}
\tablecaption{Properties of simulations used in this work.}
\tablenum{1}
\tablewidth{0pt}
\tablehead{\colhead{Simulation} & 
\colhead{$M_{200m}$} & 
\colhead{$R_{200m}$} & 
\colhead{$M_{*,90}$} & 
\colhead{$R_{*,90}$} & 
$z_{15}$ & 
$z_{2}$ & 
$z_{\mathrm{disk}}$ & 
$t_\mathrm{disk}$ & 
$d_1$ & 
\colhead{Ref} 
\\ 
\colhead{} & 
\colhead{$10^{12}\ M_\odot$} & 
\colhead{kpc} & 
\colhead{$10^{10}\ M_\odot$} & 
\colhead{kpc} & 
\colhead{} & 
\colhead{} & 
\colhead{} & 
\colhead{Gyr} & 
\colhead{kpc} & 
\colhead{} 
} 

\startdata
m12f & 1.7 & 354.7 & 6.9 & 14.28  & 3.7 & $>6.0$ & $\sim 0.61$ & $\sim 8$ & - & [1] \\
m12i & 1.2 & 314.1 & 5.5 & 9.09  & 3.1 & 3.5 & $\sim 0.34$ & $\sim 10$ & - & [2] \\
m12m & 1.6 & 341.6 & 10.0 & 12.62  & 1.9 & 2.1 & $\sim 0.34$ & $\sim 10$ & - & [3]\\
m12w & 1.1 & 300.5 & 4.8 & 8.63  & 2.7 & 3.4 & $\sim 0.34$ & $\sim 10$ & - & [4] \\
\hline
m12f (SIDM) & 1.36 & 289.8 & 6.2 & 15.7 & $\dagger$ & $\dagger$& $\dagger$  & $\dagger$ & 8.8 & [5] \\
m12i (SIDM) & 0.98 & 260 & 5.0 & 13.9 & $\dagger$& $\dagger$& $\dagger$  & $\dagger$ & 7.4 & [5] \\
m12m (SIDM) & 1.24 & 281.5 & 6.6 & 20.2 & $\dagger$& $\dagger$& $\dagger$ & $\dagger$ & 9.8 & [5] \\
\hline
m12f (DMO) & 1.28 & 284.2 & – & – & – & – & – & – & - &  [1] \\
m12i (DMO) & 0.90 & 252.8 & – & – & – & – & – & – & - & [2] \\
m12m (DMO) & 1.14 & 273.9 & – & – & – & – & – & – & - & [6]\\
\enddata



\tablecomments{All simulations have identical particle mass for dark matter ($3.5\times 10^4\ \, \rm{M_\odot}$) and initial particle mass for baryonic matter ($7.1 \times 10^3 \, \rm{M_\odot}$), as well as identical force softening parameters ($\epsilon_{\rm gas, min} = 1.0 \, {\rm pc}$, $\epsilon_{\rm star} = 4.0 \, {\rm pc}$, $\epsilon_{\rm dm} = 40 \, {\rm pc}$). $M_{200m},R_{200m}$: virial mass \& radius of the main halo at $z \sim 0$. $M_{*,90}, R_{*,90}$: 90\% of the stellar mass contained within 30 kpc of the galaxy and the spherical radius within which encloses that mass. $z_{15}$ ($z_{2}$): redshift when the cumulative fraction of stars that formed in-situ exceeded 0.5 when selecting stars at $z = 0$ within host-centric distances of 15 (2) kpc \citep{Santistevan2020}; $z_{\mathrm{disk}}, t_{\mathrm{disk}}$: redshift and cosmic time of when the disk is roughly assembled. $d_1$: the local scattering region radius which is defined as the radius at which DM particles should have experienced at least a single self-interaction \citep{vargya2021}. Assembly times for SIDM halos (i.e., values replaced by a $\dagger$) are unable to be calculated since star particle formation distances are not available for those simulations. }

\tablerefs{[1]: \citet{GK17table}; [2]: \citet{wetzel2016}; [3]: \citet{fire2}; [4]: \citet{samueltable}; [5]: \citet{omidtable}; [6]: \citet{GK19table}. }

\end{deluxetable*}

\begin{deluxetable*}{lcccccccr}
\label{tab.lmcs}
\tablecaption{Properties of identified LMC satellites used in this work.}
\tablenum{2}
\tablewidth{0pt}
\tablehead{\colhead{Simulation} & 
\colhead{$m^{\rm{peak}}_{\rm{LMC}}$} & 
\colhead{$m^{\rm{peri},*}_{\rm{LMC}}$} & 
\colhead{$m^{\rm{peri}}_{\rm{LMC}}$} & 
\colhead{$r^\mathrm{peri}$} & 
\colhead{$z^\mathrm{peri}_{\rm{LMC}}$} & 
\colhead{$t^\mathrm{peri}_{\rm{LMC}}$} & 
\colhead{$m^\mathrm{infall}_{\rm{LMC}}$} & 
\colhead{Ref} 
\\
\colhead{} & 
\colhead{$10^{10} M_\odot$} &
\colhead{$10^{10} M_\odot$} &
\colhead{$10^{10} M_\odot$} &
\colhead{kpc} &
\colhead{} &
\colhead{Gyr} &
\colhead{$10^{10} M_\odot$} &
\colhead{} 
} 

\startdata
m12f &  15.3 & 0.262 & 1.53 & 35.74  & 0.26 & 10.75 & 15.2 & [1] \\
m12i & 4.38 & 0.0388 & 1.24 & 29.53 & 0.6 & 8.04 & 3.6 & [2] \\
m12m &  3.87 & 0.0548 & 3.34 & 146.52 & 0.92 & 6.33 & 3.3 & [3]\\
m12w &  8.10 & 0.125 & 4.89 & 7.67 & 0.59 & 8.12 & 7.9 & [4] \\
\enddata

\tablecomments{$m^{\rm{peak}}_{\rm{LMC}}$: \textit{peak} bound mass of largest recent perturber. $m^{\rm{peri}}_{\rm{LMC}}$: total bound mass of largest recent perturber \textit{at pericenter}. $m^{\rm{peri,*}}_{\rm{LMC}}$: total stellar mass of largest recent perturber \textit{at pericenter}. $r^\mathrm{peri}$: distance of the perturber from the main galaxy at pericenter. $z^\mathrm{peri}_{\rm{LMC}}$: redshift when largest recent perturber reaches pericenter. $t^\mathrm{peri}_{\rm{LMC}}$: time of pericenter (13.7 Gyr = present day). $m^\mathrm{infall}_{\rm{LMC}}$: in-fall mass of the largest most recent perturber.}

\tablerefs{[1]: \citet{GK17table}; [2]: \citet{wetzel2016}; [3]: \citet{fire2}; [4]: \citet{samueltable}}

\end{deluxetable*}

\section{Methods}
\label{s.methods}

Our analysis makes use of the Latte suite of FIRE-2 cosmological zoom-in simulations \citep{wetzel2016, fire2}. These simulations capture the dynamical feedback between the live baryonic and dark matter components of MW-mass galaxies as they emerge from the cosmic web. The Latte suite consists of Milky Way-mass galaxies ($\sim 10^{12} M_{\odot}$) with different initial conditions and merger histories. We use the fiducial resolution Latte simulations with a resolution of $7100 \, \rm{M_\odot}$ per initialized gas particle and $3.5\times 10^4 \, \rm{M_{\odot}}$ per dark matter particle.

The FIRE simulations offer several important advantages in the study of halo orientations. Each galaxy is initialized with different initial conditions selected from a large cosmological box, which provides a diversity of halo assembly histories. The simulations have intervals of 20-25 Myr between snapshots, which is essential for resolving the halo response since typical dynamical times in the halo are $\sim 200$--500 Myr. The spatial resolution is sufficiently high to account for dynamical torques and the production of wakes. The feedback model implemented in FIRE leads to the global structure of the central galaxy at $z=0$ to be consistent with observational distributions \citep{bellardini22, bellardini21, sanderson+20, fire2}. We thus expect our results to be relevant for present and future studies of the shape of the Milky Way halo.

We analyze in detail the dark matter halos of four Latte galaxies (\textbf{m12f}, \textbf{m12i}, \textbf{m12m}, \textbf{m12w}) described in Table \ref{tab.sims}. These four simulations are chosen since they possess LMC-like satellites. The criterion for ``LMC-like" is defined in the following section. Additionally, two of the LMC analogs (\textbf{m12f} and \textbf{m12w}) used in this analysis were first presented in \citet{samuel21}. The visualizations\footnote{\url{http://www.tapir.caltech.edu/~sheagk/starvids.html}} of these galaxies in the Appendix illustrate the visible differences between the central galaxies due to their diverse formation histories. 

Given that LMC analogs can powerfully influence halo dynamics \citep{gc20}, we categorize and trace an ``LMC analog'' in each simulated halo by selecting the most massive and most recent/ongoing merger with pericenter distance similar to the LMC. We define LMC-like satellites as simulated dwarf galaxies that meet the following criteria: (a) Reaches at least a total mass of $4\times10^{10} \, \rm{M_\odot}$ or a stellar mass of $5\times10^8 \, \rm{M_\odot}$, (b) the satellite crosses into the virial radius of the host, (c) is not destroyed after 6 Gyr after the simulation begins. We refer to the interaction as a ``major merger'' if the LMC analog has an infall mass-to-central galaxy bound mass ratio of $M_{\rm{LMC}} / M_{\rm{MW}} > 1/3$ and as a ``minor merger'' otherwise. 

All four galaxies have either a major or minor merger with a simulated LMC analog across their evolutionary histories. These simulated mergers allow us to test the effect of an LMC-like merger on the DM symmetry axes as a function of redshift and mass ratio:
\begin{itemize}
    \item \textbf{m12i} is a relatively stable galaxy with no recent major mergers, which makes it an excellent baseline condition at $z=0$ for comparison to simulations with recent major mergers. It has an LMC analog that has a minor merger with the host at $z=0.6$.
    \item \textbf{m12f} has a recent major merger ($M_{LMC} \geq \frac{1}{3}M_{MW}$) involving an LMC-like satellite with stellar mass of $\sim10^9 \, \rm{M_\odot}$. 
    \item \textbf{m12w} has a major merger with an LMC companion but has few stellar streams at present. 
    \item \textbf{m12m} is a galaxy with a minor LMC merger ($M_{LMC} < \frac{1}{3}M_{MW}$) at relatively early times. It also has the most massive stellar disk, which forms very early in the simulation. 
    \end{itemize}

\subsection{Determining dark matter symmetry axes}
\label{ss.symmaxes}

\begin{figure}[htbp]
    \centering
    \includegraphics[width=.5\textwidth]{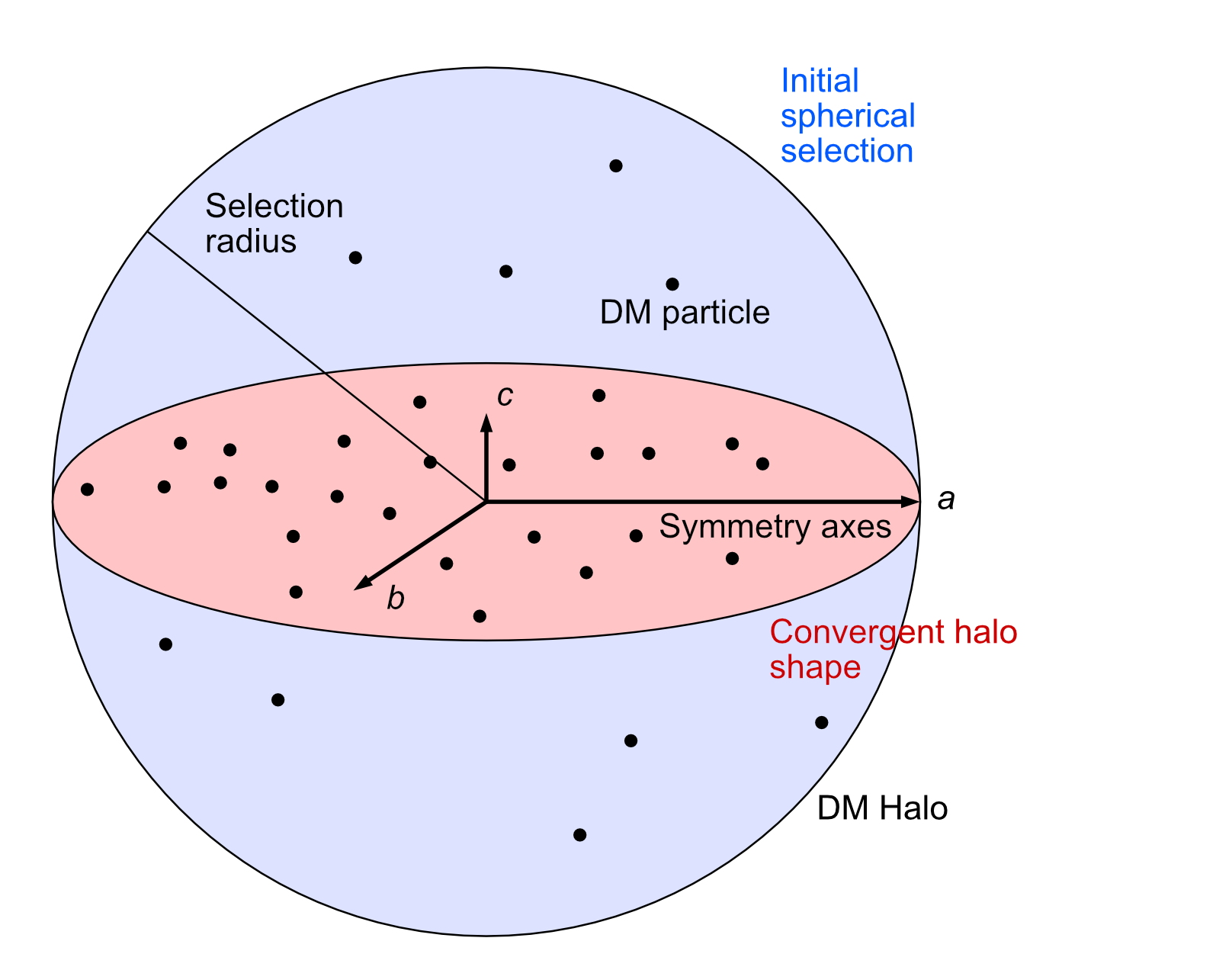}
    \caption{The selection radius defines the enclosed spherical region where dark matter particles are chosen. The moment of inertia tensor is determined from the enclosed dark matter particles. Symmetry axes of the dark matter halo (labeled $a, b, c$) are defined by the eigenvectors of the reduced moment of inertia tensor. The length of each symmetry axis is the square root of the corresponding eigenvalues. These eigenvectors and eigenvalues define the converged halo shape.}
    \label{fig:dm_axes}
\end{figure}

 Figure \ref{fig:dm_axes} illustrates the process of determining the DM symmetry axes. We find the symmetry axes of the dark matter halo by diagonalizing the reduced moment of inertia tensor:

\begin{equation}
    \label{moi}
    I_{ij} = \Sigma_{u_k \in V} \frac{u_{i,k} u_{j,k}}{d_k^2}
\end{equation}

where $V$ is the set of position vectors $\vec{u}$ of each dark matter particle, and

\begin{equation}
    d_k = \sqrt{u_{1,k}^2 + \bigg(\frac{u_{2,k}}{q}\bigg)^2 + \bigg(\frac{u_{3,k}}{s}\bigg)^2}
\end{equation}
where $q = b/a$ (ratio of the intermediate axis to major axis lengths) and $s = c/a$ (ratio of the minor axis to the major axis). We initialize $q = s = 1$ for the sake of simplicity, assuming that the axis orientation is independent of the axis ratios \citep{vargya2021, allgood2006}.

We calculate and diagonalize the reduced inertia tensor for each galaxy at different selection radii (i.e., selecting DM particles from a sphere of greater volume with each iteration). The eigenvectors from the resulting diagonalization are the symmetry axes ($\vec{u}, \vec{v}, \vec{w}$, for major, median, and minor axes) for the dark matter halo at that particular selection radius. The minor axis of the dark matter halo is defined to be the eigenvector with the smallest corresponding eigenvalue. The axis lengths of each symmetry axis (eigenvector) are defined to be the square root of each eigenvalue.

We emphasize that the particles are selected within ellipsoidal volumes rather than shells. Contrarily, \citet{zemp11} find that the best method to calculate local matter distributions is by using ellipsoidal shells without distance-based weighting and omitting subhalo particles. We opt for volume selection since the axes fitted to the particles within that volume can inform the force that stellar tracers are sensitive to. In this work, we opt to keep the subhalos within our selection sample. Simplified potential models of the MW do not attempt to model the influence of individual subhalos. To this end, this work is mainly geared towards detecting mismatches between this simplified model and a more realistic model due to the presence of massive subhalos.

\subsection{Calculating the Obliquity Angles with Host Axes with Degeneracy Corrections}
\label{ss.obliq}

We define the obliquity angle of each simulated Milky Way as the angle between the DM minor axis, determined as described in section \ref{ss.symmaxes}, and the short symmetry axis of the disk of the central galaxy which is calculated in a similar manner to the DM using the youngest 25\% of in-situ stars within a radius that contains 90\% of the mass selected within $10$ kpc of the halo center. The obliquity angle is then determined from the dot product between the minor axis and the DM minor axis at a particular galactic radius and/or snapshot in time: 
\begin{equation}
    \theta = \arccos(w_{gal} \cdot w_{DM})
\end{equation}

The process of diagonalizing the moment of inertia tensor only determines the symmetry axes up to the sign of the eigenvectors,  leading to a degeneracy that we need to account for. For degeneracy corrections, we choose the stellar disk angular momentum within $r=10$ kpc at present day as a reference vector.

To break these degeneracies, we invert the major (axis vector corresponding to length a) and minor (axis vector corresponding to length c) symmetry axes if the dot product between the DM minor axis and the disk angular momentum vector (the median angular momentum of the stellar disk within $r=10$ kpc) is negative.

The angular momentum vector is defined as:

\begin{equation}
    L_{\rm{disk}} = \sum\limits_{r<10\rm{kpc}}m_i (\vec{r_i} \times \vec{v_i})
\end{equation}

and the normalized angular momentum vector is:

\begin{equation}
    \hat{L}_{\rm{disk}} = L_{\rm{disk}} / ||\sum\limits_{r<10 \, \rm{kpc}} m_i (\vec{r_i} \times \vec{v_i})||
\end{equation}

For consistency, the minor axis is always defined as the cross product of the calculated major and median axis eigenvectors such that our rotation tensor obeys the right-hand rule. 

In Figure \ref{fig.m12i_density}, the direction of the DM symmetry axes (colored circles), the symmetry axes (black), and the angular momentum vector (red) described in section \ref{ss.obliq} are overlaid on the DM density map of \textbf{m12i}. The orientations of the dark matter halo at $z=0$ are calculated based on the angle between the DM symmetry axes and the host symmetry axes as shown in the figure.

For analyzing the orientation as a function of redshift, we select DM particles within initial spheres of radii of $r=R_{*,90}, 5R_{*,90}, R_{\rm 200m}$, where $R_{*,90}$ is the radius within which 90 percent of the star particles are contained and $R_{\rm 200m}$ is the virial radius for each particular galaxy at a given redshift. We calculate the orientation relative to the galaxy axes at that particular redshift as described in \ref{ss.symmaxes}, with the short axis defining the orientation of the positive z-axis.

Additionally, we make note of the triaxiality as a function of redshift for each initial volume (see Section \ref{ss.triax}) as this can determine the axis of rotation of the symmetry axes. We caution that the shape measurements may be biased due to the presence of substructure as mentioned in Section \ref{ss.symmaxes}, which may dominate the shape measurement at the radius of the subhalos.

\subsection{SIDM and DMO Simulations}
\label{ss.SIDM_DMO}
Latte galaxies \textbf{m12f}, \textbf{m12i}, and \textbf{m12m} have additional simulations that are initialized with self-interaction dark matter with baryons (SIDM) and cold dark matter without baryons (DMO) \citep{omidtable, GK19table, GK17table, wetzel2016}. We replicate the orientation calculation for the SIDM variants identical to the process aforementioned. However, this process needs to be tweaked for DMO simulations given there is no stellar (baryonic) disk to serve as a reference frame. We define a pseudo-stellar disk by calculating the moment-of-inertia tensor at $r=10 $ kpc for each halo and using the short axis of the reduced tensor as the "stellar disk" reference vector. We caution that the orientation of the pseudo-stellar disk, i.e. the inner DM halo itself, would reflect stronger alignments to the outer halo since the inner halo is no longer coupled to a distinct stellar disk.

\subsection{Triaxiality and Determining Effect of LMC-like Mergers}
\label{ss.triax}

We calculate the triaxiality (shape parameter) of the halo at the snapshot of interest. The triaxiality is defined as a function of the axis lengths $(a,b,c)$ that correspond to the major, median, and minor axis: 

\begin{equation}
     T = \frac{a^2 - b^2}{a^2 - c^2}
\end{equation}

We consider a halo to be prolate if $1\geq T>2/3$, triaxial if $2/3>T>1/3$, and oblate if $T<1/3$ \citep{vargya2021, prada2019, allgood2006,  vera2011}. We hypothesize that the axis of rotation should be roughly orthogonal to the LMC position. Transitively, one of the axes orthogonal to the rotation axis should closely align to the LMC analog position as we expect the perturber to shift the DM particle density in either the median or major axes. We determine the alignment to the major axes across all our galaxies in our sample regardless of triaxiality (shape), however, the shape of the halo can be useful in probing perturbations of the halo.

To constrain the triaxiality accurately, we choose $(a,b,c)$ on an iterative code that deforms the ellipsoid for a maximum of 100 iterations as implemented in \cite{vargya2021} at each snapshot in time (at different lengths). The iterative code optimizes the shape of the ellipsoid by recursively reshaping the selection geometry to the previously estimated ellipsoid shape. This process is repeated until either the maximum iteration count has passed, or until the ratios of the axis lengths converge where $(a, b, c)$ are the axis lengths.

This method is identical to the method described in Section 2.2, except for the triaxialities being determined by iterating over 100 selection ellipsoids rather than the single iteration of the code used to determine the symmetry axes. We find that iterating the selection ellipsoid leads to a better-constrained halo shape at the cost of poorly constrained obliquities in toy models.

The alignment of the DM halo to the simulated LMC satellite ($\theta_{LMC}$) is determined by the alignment of the major axis ($\vec{a}$) and the position of the satellite ($\vec{x_{LMC}}$) relative to the host. The symmetry axis degeneracy is broken by choosing the axis vector that minimizes the angle between the LMC and the axis vector. These major axes are derived from the reduced inertia tensors of the DM particles at different scale lengths ($R_{*,90}$, $5R_{*,90}$, $R_{\rm 200m}$).

\begin{equation}
    \theta_{LMC} =\min( \cos^{-1}(\vec{x}_{LMC} \cdot \vec{a}, \vec{x}_{LMC} \cdot (-\vec{a})))
\end{equation}

\subsection{Settling of the Galaxy Orientation and Formation of the Galactic Thin Disk}
\label{ss.disk_settling_methods}

Given the moment of inertia of the stellar component of the galaxy as a reference frame for calculating the halo position angle, it is of interest to know when the thin stellar disk begins to form and if it forms before or after the stellar component vector becomes influenced by the orbital vector of the LMC analog. The coherence of the thin disk appears when the star-forming HI gas becomes rotationally supported within the galaxy. We quantify the rotational support strength of the cold HI gas as $\langle v_\phi \rangle / \sigma_{v,\phi}$. This ratio describes the rotational support of gas particles within the inner galaxy ($r < 0.05\mathrm{R_{vir}}$) as the mean rotational motion of the cold gas against the random azimuthal motions. We define the time at which the thin disk has or begins to appear coherent as $t_{\rm disk}$ which is determined by the transition time when roughly the rotational support ratio attains half the maximum mean rotational velocity of the cold gas.

\section{Results and Discussion}
\label{s.ResultsAnddiscussion}

\subsection{Halo Orientation and Shape at $z=0$}
\label{ss.obliq_z0}

\begin{figure*}[htbp]
    \centering
    \includegraphics[width=.9\textwidth]{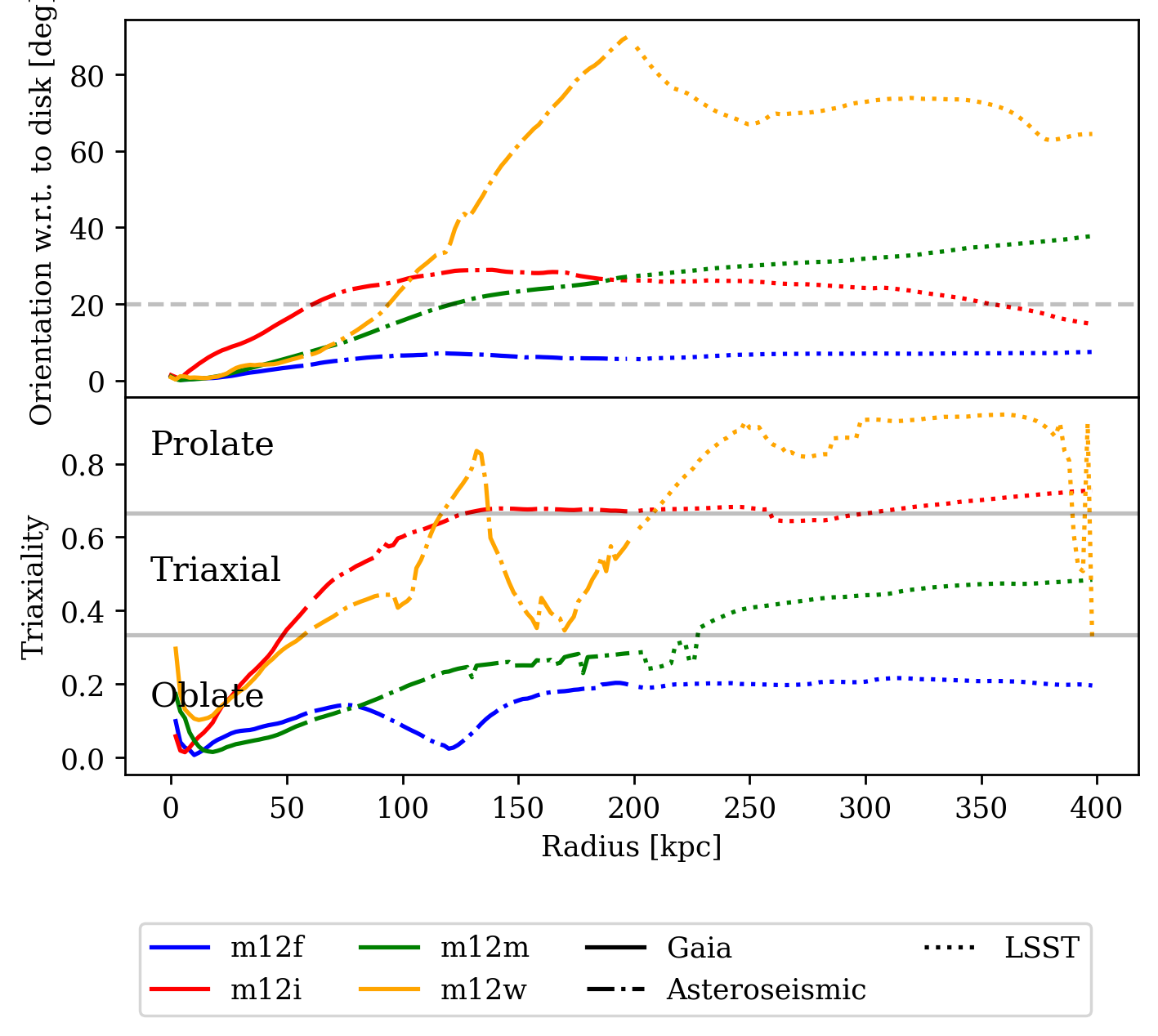}
     \caption{\textbf{Upper panel}: DM-host minor axis orientation plots of \textbf{m12f}, \textbf{m12i}, \textbf{m12m}, and \textbf{m12w} as a function of radius at $z=0$. The distance ranges for Gaia (for MSTO), ground-based M-giant asteroseismology (AS), and LSST are shown with different line styles. \textbf{Lower panel}: DM triaxiality plots of \textbf{m12f}, \textbf{m12i}, \textbf{m12m}, and \textbf{m12w} as a function of radius at a redshift of $z=0$. Triaxiality is determined by the convergent ellipsoid at each radii. The horizontal lines demarcate the transitions between oblate ($T<1/3$), triaxial ($1/3<T<2/3$), and prolate ($T>2/3$) shape classifications. This plot is available as a function of time in Figure \ref{fig.m12_triax_z}.
    }
    \label{fig.m12_triax_orientation_z0}
\end{figure*}

In the upper panel of Figure \ref{fig.m12_triax_orientation_z0}, the dark matter halo symmetry axes have greater obliquities at greater radii. We note that \textbf{m12f} is more highly aligned to the stellar disk as its orientation at $r=100$ kpc is $\theta<10\deg$ compared to the $\theta \sim 20\deg$ orientation of the other galaxies, possibly due to the merger happening edge-on to the stellar disk. Most notably, we find that \textbf{m12w} diverges in orientation significantly at $r \geq 100$ kpc. This may be attributable to the aforementioned massive satellites that are in-falling within 100-200 kpc of the halo center. Overall, these findings are consistent with the tentative evidence of a radially-varying halo orientation in the Milky Way as found in \cite{vas21}.

To quantify at what particular radius the halo diverges, we calculate the fraction of the virial radius at which a particular halo tilts more than 20 degrees relative to the disk. Halos that meet this divergence criterion diverge at about 0.2-0.4 $R_{\rm 200m}$ (Table \ref{tab.divergence}). In comparison to a study of halo orientations (relative to the angular momentum of the stellar disk), \citet{prada2019} find that a plurality of the simulated Aquarius halos aligns well to the stellar disk at all radii (most not exceeding ~20-30 degrees in obliquity). \citet{prada2019} also find that some halos experience extreme twisting (some near perpendicular to the disk even at $0.5R_{\rm 200m}$). Although our analysis makes use of a much smaller halo sample, we observe similar trends—\textbf{m12f} is a prime example of a well-aligned halo at all radii and \textbf{m12w} is a prime example of extreme twisting.

The lower panel of Figure \ref{fig.m12_triax_orientation_z0} shows the triaxiality of the DM halo as a function of radius at $z=0$ for the four galaxies. \textbf{m12f} and \textbf{m12m} are either oblate or triaxial within 100 kpc, while \textbf{m12i} and \textbf{m12w} extend into prolate configurations further from the galactic center. In particular, between 100-150 kpc, \textbf{m12f} experiences a dip in triaxiality. Further investigation reveals the presence of an in-falling satellite within that range with a bound mass of $1.4\times 10^{11} \, \rm{M_\odot}$. Similarly, \textbf{m12w} has a triaxiality shift between 100-200 due to several satellites within the vicinity (the most massive having a bound mass of $8.1\times 10^{10} \, \rm{M_\odot}$). The shapes of halos are possibly biased by the presence of these subhalos, and may not reflect the true shape of the local mass distribution.
Furthermore, the inner halos of all these galaxies are oblate, which is consistent with the distribution of the inner DM halo being influenced by the baryonic disk. However, recent phase space analyses of the Helmi stream have constrained a roughly prolate inner halo potential \citep{dodd2022}. They acknowledge that the potential model used in their analysis ignores possible dynamical influences from massive satellites, e.g., introducing reflex motion of the MW towards Sagittarius \citep{vas21}.

\begin{figure*}
    \includegraphics[width=1\textwidth]{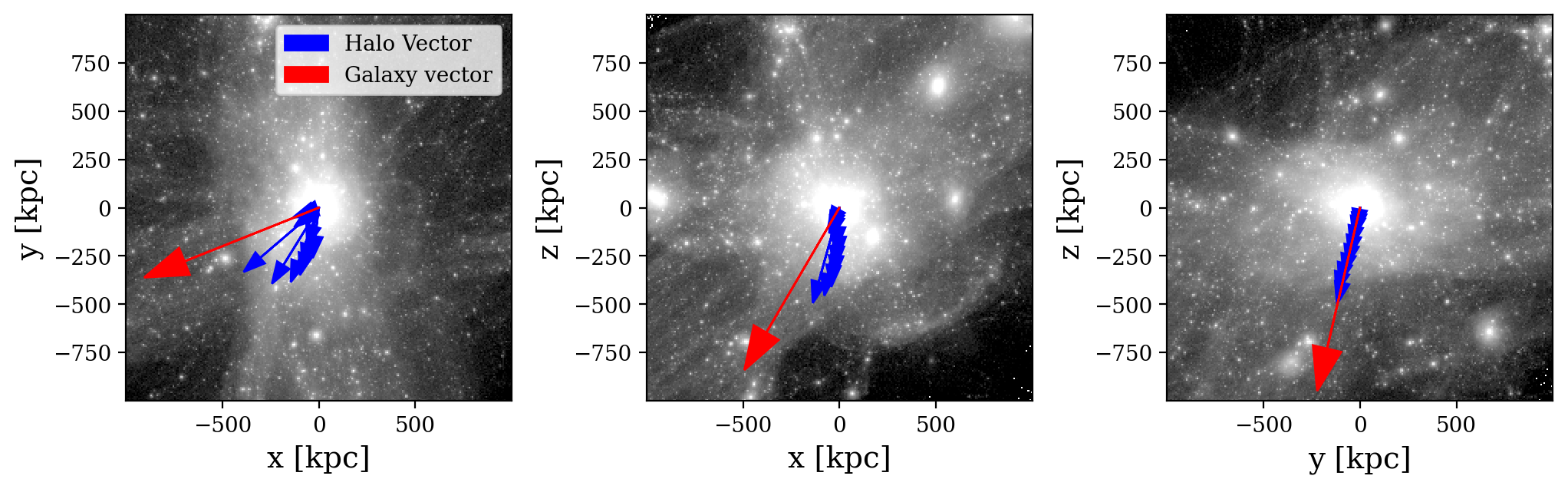}
    
    \caption{DM density plot of \textbf{m12i} with host minor axes over-plotted at $z=0$. The red arrow indicates the direction of the stellar disk/galaxy vector. The blue arrows indicate the minor axis of the halo, with the length scaled to the radius at which the principal axes were calculated. Each column shows a different projection of the halo. Brighter regions indicate higher densities of dark matter. Figure \ref{fig:m12_allaxes} replicates this density map for all galaxies.}
    \label{fig.m12i_density}
\end{figure*}

\setcounter{table}{2}
\begin{table*}[htbp]
    \centering
    
    \begin{tabular}{|l|l|l|l|}
        \hline
        \hline
        Simulation & $r_{\rm div}$ {[}kpc{]} & $r_{\rm div} / R_{\rm 200m}$ & $r_{\rm div} / R_{*, 90}$ \\
        \hline
        m12f & -- & -- & --\\
        m12i & 64 & 0.2 & 7.0\\
        m12m & 124 & 0.36 & 9.8\\
        m12w & 96 & 0.32 & 11.1\\
        m12m (SIDM) & 230 & 0.82 & 10.0\\
        \hline
        \hline
    \end{tabular}
    \caption{Approximate radii at which each Latte galaxy's halo orientation exceeds ~20 degrees (divergence radius $r_{\rm div}$). Divergence also shown as a fraction of the virial radius ($r_{\rm div} / R_{\rm 200m}$) and disk size ($R_{*,90}$). The only SIDM/DMO halo that diverged was \textbf{m12m} (SIDM).}
    \label{tab.divergence}
\end{table*}

\begin{figure*}[htbp]
    \centering
    \includegraphics[width=.85\textwidth]{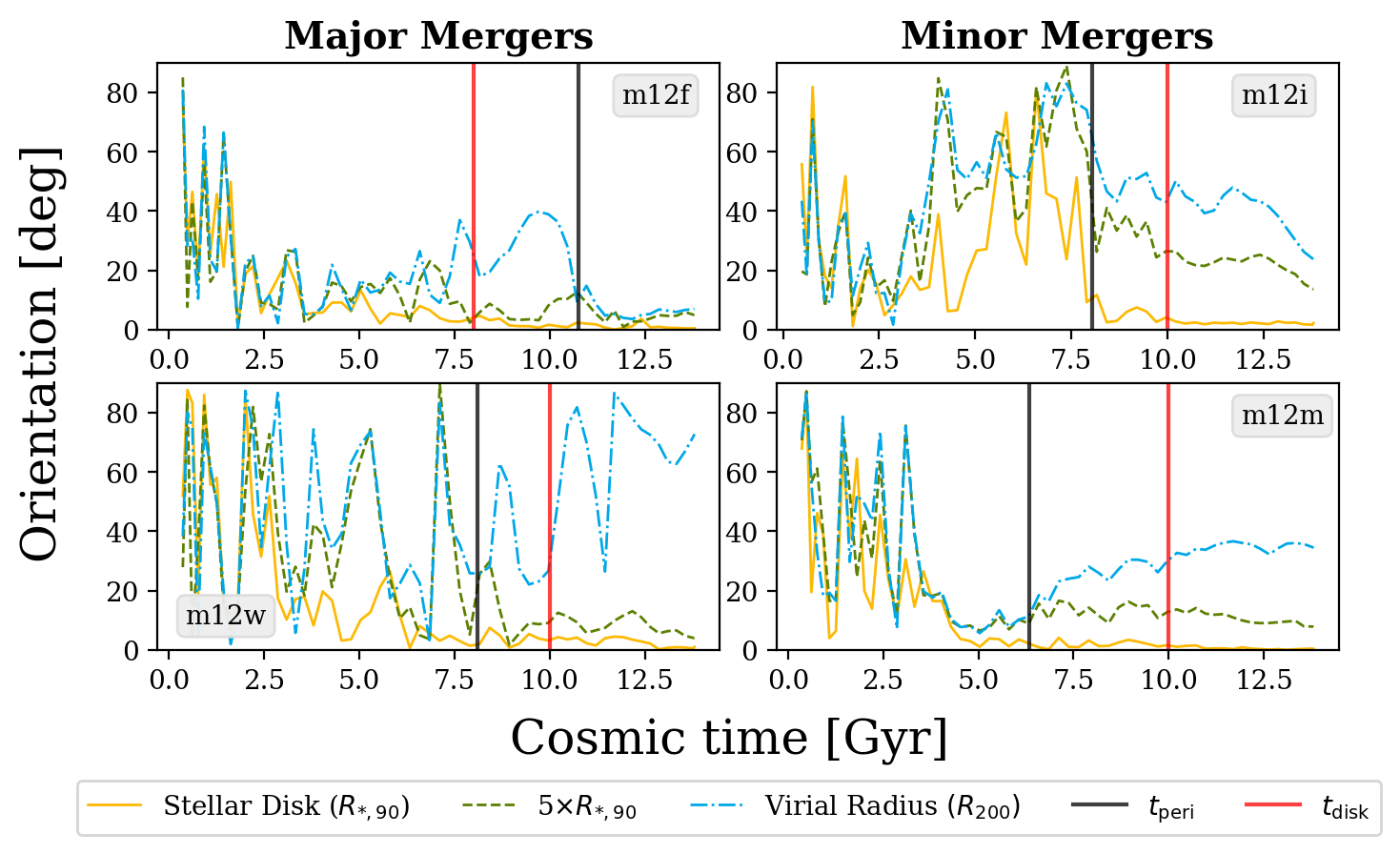}
    \caption{Change in the orientation of the DM halo over time at different scale lengths. The yellow solid line is the orientation of the DM halo at the edge of the stellar component of the galaxy ($R_{*, 90}$), the green dashed line is the orientation outside the galaxy but well within the DM halo ($5 \times R_{*,90}$), and the blue dashed-dotted line is the global orientation of the DM halo at $R_{\rm 200m}$. Orientations at each snapshot are calculated with respect to the galaxy vector at that particular snapshot. The vertical black line indicates when the LMC analog reaches pericenter for the particular galaxy and the red line indicates when the thin disk of the galaxy begins to form.} 
    \label{fig.m12_orientations_z}
\end{figure*}

\begin{figure*}[htbp]
    \centering
     \caption{Same as Figure \ref{fig.m12_orientations_z}, however orientations at each snapshot are calculated with respect to the stellar disk \textbf{at present day}.}
     \includegraphics[width=.85\textwidth]{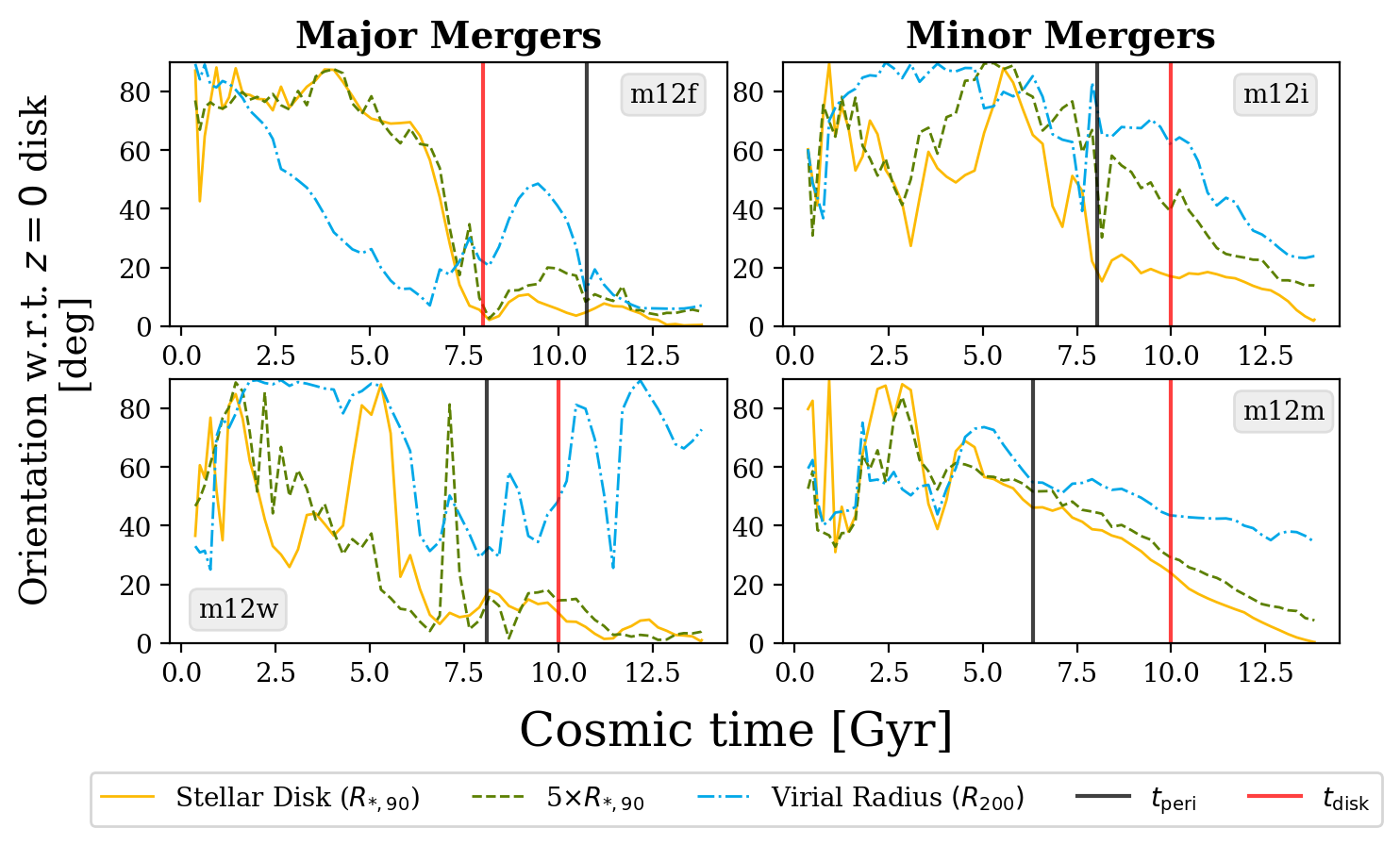}
    \label{fig.m12_orientations_z_at_z0}
\end{figure*}

\subsection{Halo Orientations Across Redshift}
\label{ss.obliq_across_z}

In Figure \ref{fig.m12_orientations_z}, we show the orientations of the DM halo over time with respect to the evolving galaxy axes. We observe central radii aligning well to the galaxy symmetry axes as the galaxies approach $z=0$. Furthermore, at the virial radius, we see the minor axes are less aligned to the galaxy as the simulation approach $z=0$.

In the outer halo ($r = 5R_{*,90}, R_{\rm 200m}$), we observe the halos with a major merger quickly align towards the disk as the LMC analog reaches pericenter with the halo. While the trend is present in the major-merging halo, the minor-merging halos have relatively smooth transitions.

The outer halo does not shift its density to be consistent with the inner halo; however, the inner halo orients the disk towards the LMC analog as it approaches pericenter. In Figure \ref{fig.m12_orientations_z_at_z0}, we show the time-evolving orientations of the halo relative to the present-day axes. At around pericenter, we observe that the inner halo aligns to the present-day disk within 1-2 Gyr of pericenter of a major-merging satellite. This follows from the quick response of the inner halo to dynamical perturbations. In our CDM sample, the average dynamical timescale within the present-day divergence radius is roughly 2-4 Gyr, which is fast enough to allow for the inner halo/disk to respond to the LMC analog. Although this response is noticeable in minor-merging \textbf{m12i}, it is comparably weaker than the more massive satellite mergers.

These results suggest that massive LMCs ($\sim10^{11} \, \rm{M_\odot}$) play a critical role in determining the orientation of the Milky Way's disk at present day. This result is consistent with the findings in \cite{santistevan21}, where the orientation of the disk is heavily influenced by the most recent massive gas-rich merger.

\subsection{SIDM and DMO Comparisons}
\label{ss.sidm_dmo_results}

\begin{figure*}[htbp]
    \centering
    \includegraphics[width=.9\textwidth]{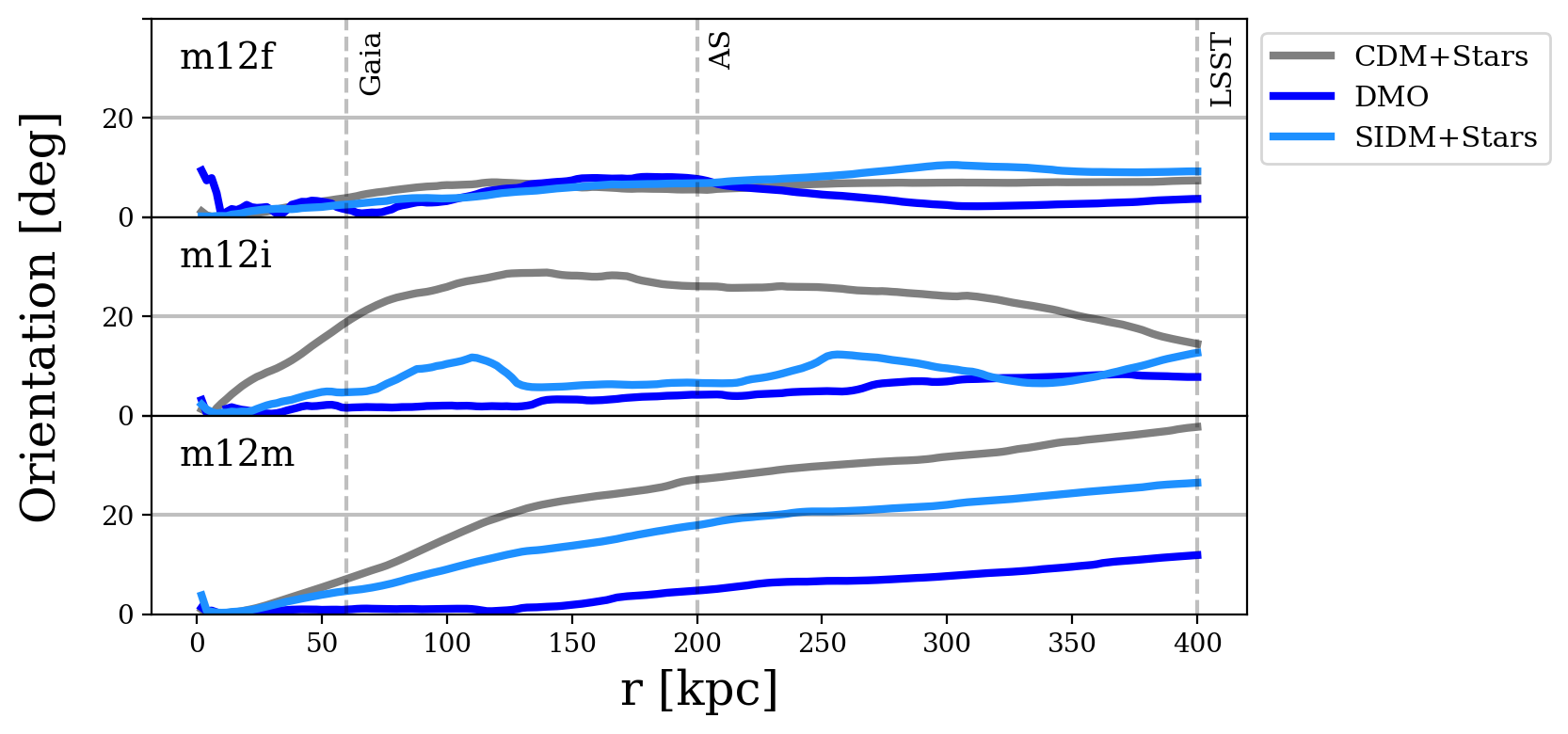}
    \caption{Halo orientations relative to the stellar disk for different simulations at $z=0$. Each galaxy begins with identical initial conditions with the exception of how dark matter interacts within the system. The gray line represents the orientation of the halo with CDM and baryons (see Figure \ref{fig.m12_triax_orientation_z0}), and the light-blue (dark-blue) lines correspond to DMO (SIDM) halo orientations. Additionally, we plot distance ranges for Gaia (MSTO), M-giant asteroseismology (AS), and LSST. The DMO pseudo-stellar disk axis is defined to be the angular momentum of DM particles within 10 kpc.} 
    \label{fig.m12_orientations_z0_diff_species}
\end{figure*}

We extend our analysis to three available Latte galaxies (\textbf{m12i, m12f, m12m}) initialized with SIDM with baryons and CDM-only simulations. Our results, shown in Figure \ref{fig.m12_orientations_z0_diff_species}, plot the present-day orientations for these different simulations. Compared to the original orientations at present day (see Figure \ref{fig.m12_triax_orientation_z0}), the SIDM and CDM-only halos in \textbf{m12i} and \textbf{m12m} have smaller halo-galaxy obliquities as a function of radius. Furthermore, we find that the only halo that diverges ($>20$ degrees) is the SIDM halo in \textbf{m12m} at ~0.7 $R_{\rm 200m}$. In \textbf{m12m}, we see that the orientations between CDM, DMO, and SIDM are similar to each other. In \citet{prada2019}, they find that DMO orientations are weakly correlated with each other at all radii. This finding appears to be consistent when comparing the DMO and CDM orientations. More insight can be gained from future analysis of the correlation between the SIDM and DMO orientations.

In our analysis of SIDM galaxies, we make note of the orientations calculated within $d_1$ (the local self-interaction scattering region). This particular halo region is of import given the physical differences between SIDM and CDM halos become significant due to SIDM particles having experienced at least one self-interaction. However, we observe that the differences in orientation between CDM and SIDM halos within $d_1$ are not apparent. This result indicates that the effects of self-interaction do not impact the orientation of the halo relative to the disk at small scales.

Given that the elastic SIDM orientations appear to be oriented closer to the disk at all radii, we expect this effect to be even stronger in alternative SIDM scenarios such as dissipative SIDM (dSIDM) model \citep{shen21}. Within the dSIDM model, DM can form dark disks, or inner DM halos that behave as oblate rotators, that can more strongly force alignment with the baryonic disk \citep{shen22, shen21}.

\subsection{Orientation of DM Halo Relative to LMC Infall}
\label{ss.obliq_across_lmc}

\begin{figure*}[htbp]
    \centering
    \includegraphics[width=.85\textwidth]{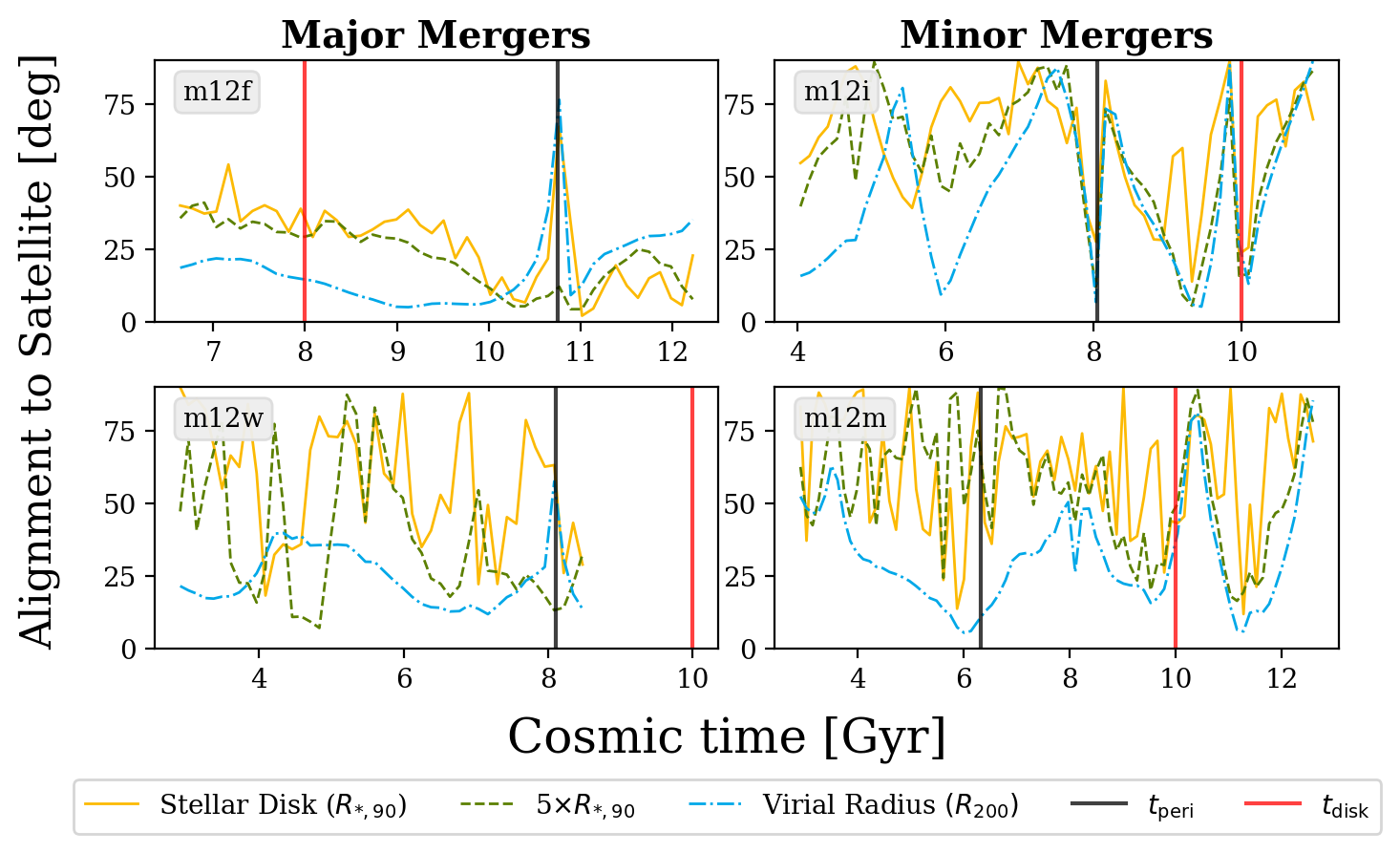}
    \caption{
    Angle between the major axis to the LMC analog position. The black line indicates the time when the satellite reaches pericenter and the red line indicates when $t_{\rm bursty}$ occurs.} 
    \label{fig.m12_lmc}
\end{figure*}

\begin{figure*}[htbp]
    \centering
    \includegraphics[width=.85\textwidth]{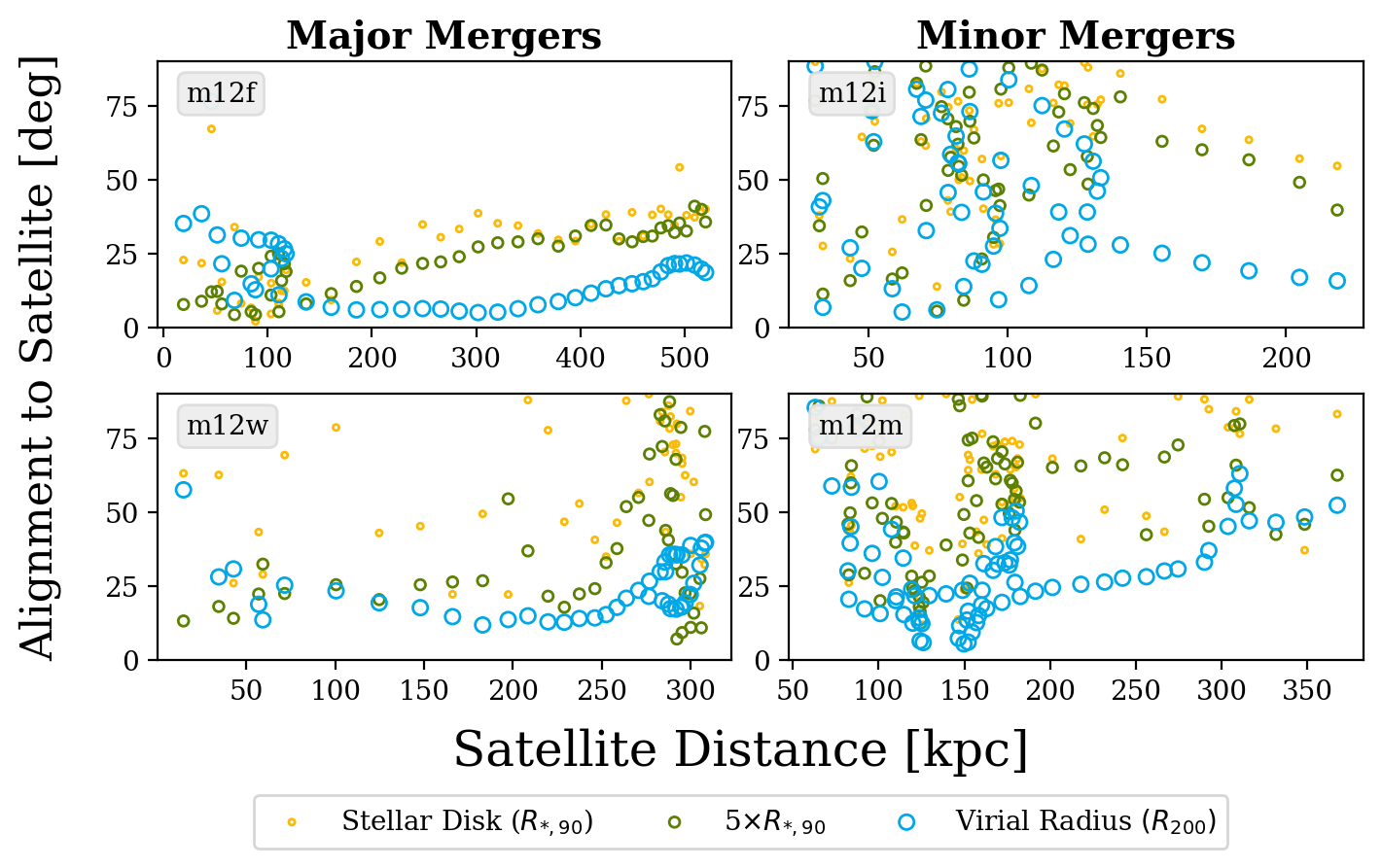}
    \caption{Major axis alignment to LMC as a function of distance to host. Different selection scale lengths for the DM halo are coded by color. The virial radius tends to be more closely aligned with the LMC than the inner halo, showing differential alignment to the LMC over time.} 
    \label{fig.align_r}
\end{figure*}

In Figure \ref{fig.m12_lmc}, we show the orientations of the DM halo relative to the LMC position as a function of time at different radii. We analyze all four CDM+Baryon galaxies with different LMC analog mergers. We indicate the time of the LMC pericenter, total peak mass, and different present-day scale lengths are shown in Table \ref{tab.lmcs}. The major-merging ($M_{\rm{LMC}} > \frac{1}{3}M_{\rm{MW}}$) galaxies are \textbf{m12f} and \textbf{m12w}, while the minor-merging ($M_{\rm{LMC}} < \frac{1}{3}M_{\rm{MW}}$) galaxies are \textbf{m12i} and \textbf{m12m}. Additionally, in Figure \ref{fig.align_r}, we present these orientations as a function of distance to the central galaxy. 

We observe the major merger in \textbf{m12f} corresponds to increased major axis alignment to the LMC position over time, with a sudden period of misalignment when the LMC reaches its pericenter (at $r=36$ kpc). In \textbf{m12w}, we see the major axis of the galaxy aligning well over time as the satellite approaches pericenter, with a short burst of misalignment following pericenter. The misalignment may be attributed to the shifting center of mass of the central galaxy as the LMC passes pericenter. Across both major-merging galaxies, their outer halos are strongly aligned to their respective LMC analogs (Figure \ref{fig.align_r}).

In our minor-merging galaxies, \textbf{m12i} and \textbf{m12m} show a weak correlation between the halo orientation and direction of the merger as a function of time and distance (Figure \ref{fig.align_r}). We hypothesize that due to the lower mass LMC, the satellite applies less force on the host axes and does not influence the axes significantly. 

Overall, the orientations of the outer halo in Figure \ref{fig.m12_lmc} appear consistent with major axis (a density dipole) aligning and tracking the LMC over time, but the mass must be $\geq \frac{1}{3}M_{\rm{MW}}$ for this to happen. This is consistent with \cite{vas21}, where they find that the outer halo elongates towards the previous position of the LMC a few hundred Myr ago in fitted static Milky Way potentials.

Across both minor and major merging galaxy samples, we observe that the outer halo tends to be more well-aligned to the in-falling satellite position than the inner halo. These results are consistent with how simulated LMCs can effectively decouple the inner and outer DM halo per the findings of \cite{dmwake}. These results are similar to the findings in \cite{shao2021} where they explored the twisting of the DM halo in MW-like galaxies in EAGLE simulations. They find the outer halo (at $R_{\rm 200m}$) is more aligned to the orbital plane of satellites, and that the presence of an LMC-mass satellite does not affect the orientation of the outer halo itself but can decouple the orientation of the central disk relative to the outer halo \citep{shao2021}. However, they find that they cannot constrain a typical scale length (``twist radius'') at which the inner and outer halo diverge and find this transition radius can vary between 30-150 kpc \citep{shao2021}.

\begin{figure*}[htbp]
    \centering
    \includegraphics[width=.8\textwidth]{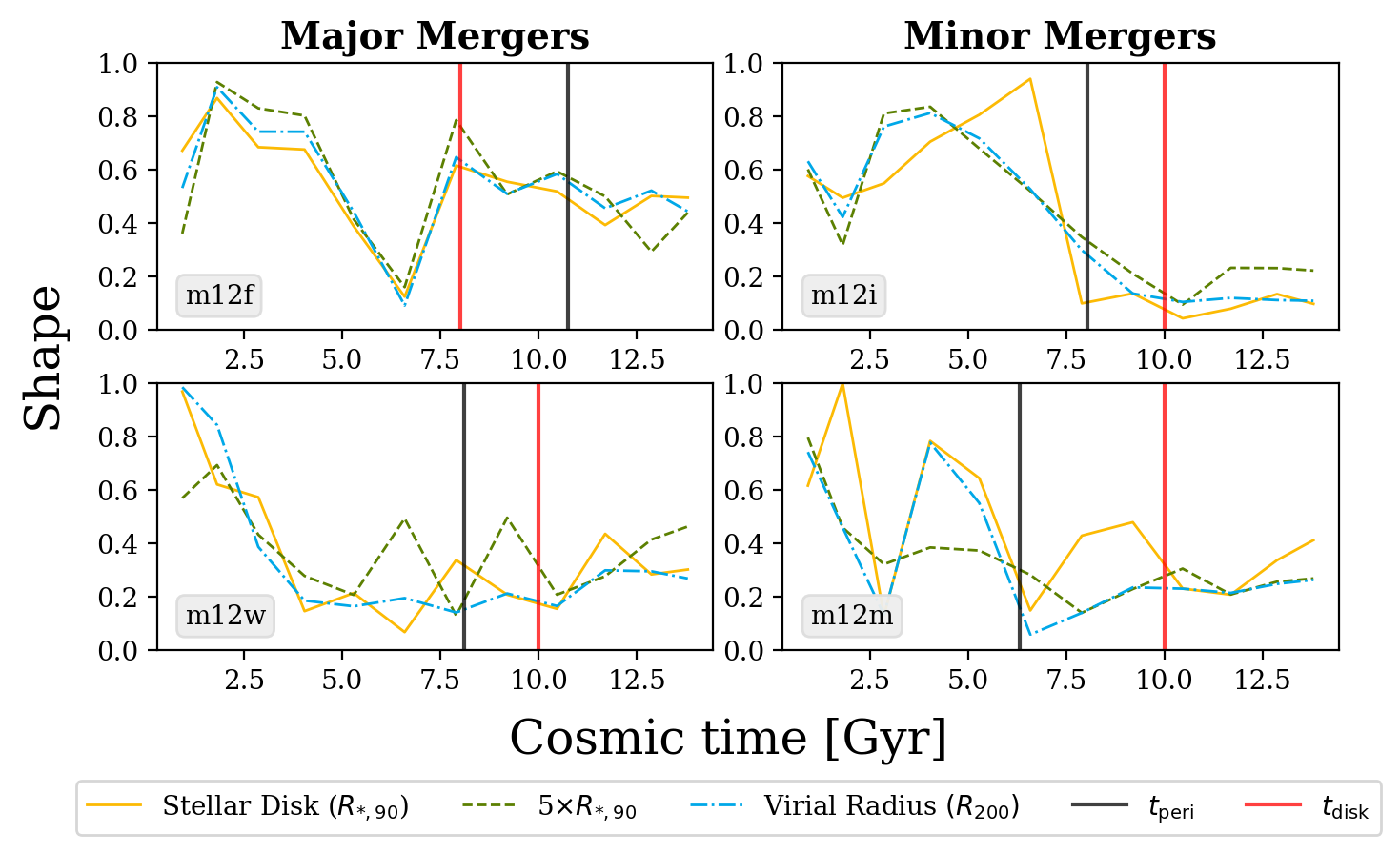}
    \caption{Halo triaxiality as a function of time at different scale lengths. The LMC pericenters of each galaxy is indicated with arrows.} 
    \label{fig.m12_triax_z}
\end{figure*}

In Figure \ref{fig.m12_triax_z} we present the triaxiality as a function of time. The triaxialities for major-merging galaxies appear to be more erratic over time. On the other hand, minor-merging galaxies have a more gradual triaxiality transition as a function of $z$. The DM halo of \textbf{m12i} appears to have a strong change in shape as the LMC reaches pericenter.

\begin{figure*}[htbp]
    \centering
    \includegraphics[width=.9\textwidth]{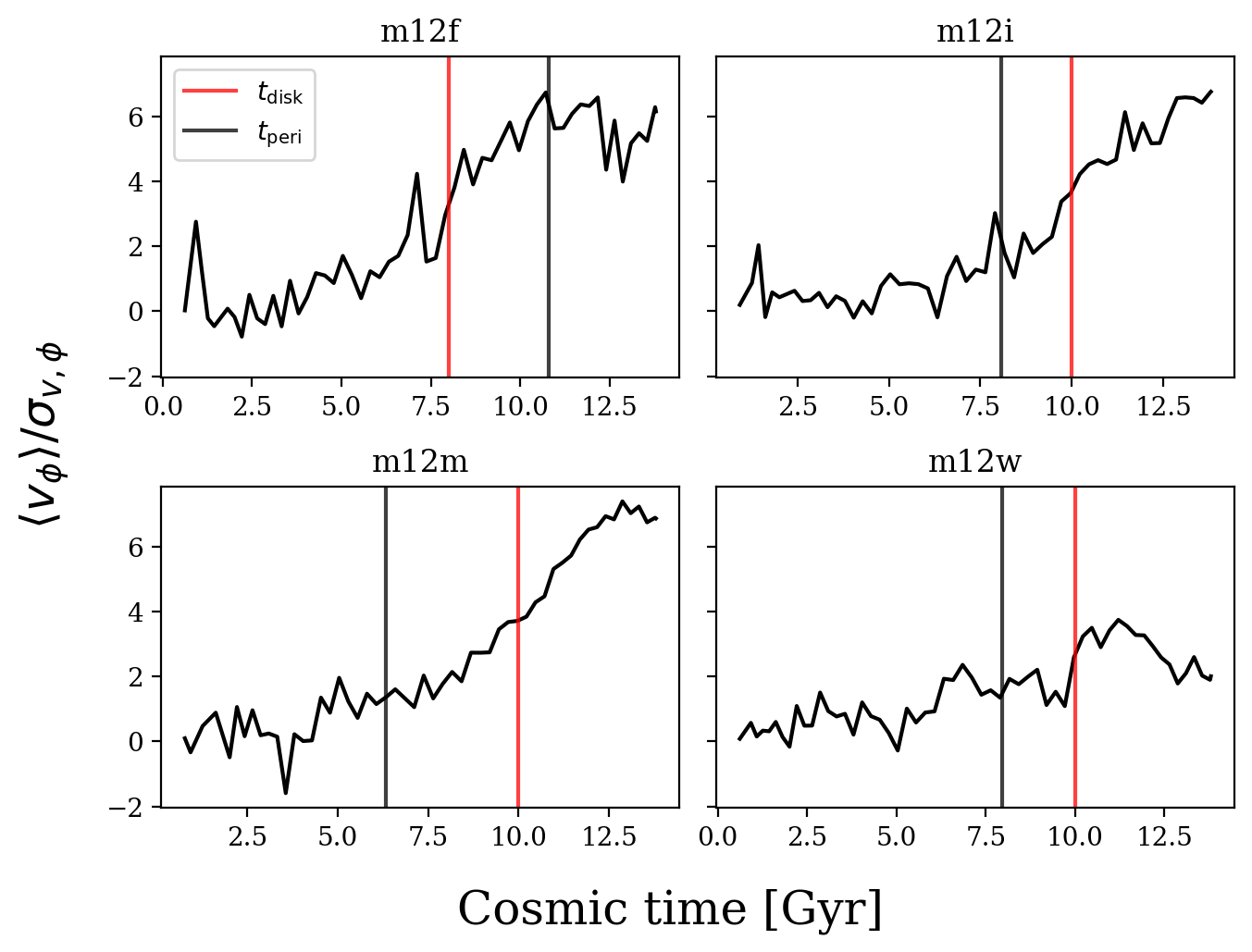}
    \caption{$\langle v_\phi \rangle / \sigma_{v, \phi}$ ratios for CDM+Baryon galaxies. Vertical red line indicates the time when the thin disk becomes roughly assembled. Verticle black line indicates the time of LMC-analog pericenter. The reference frame is defined as the cylindrical principal axes where the $z$–direction is defined by the minor axis of the moment of inertia of young stellar particles.} 
    \label{fig.m12cdm_vsr}
\end{figure*}

\subsection{Orientation of the Halo Short Axis within Filaments}
\label{ss.short_axis_filaments}

At present day, all galaxies in our analysis have their minor axes in the outer halo aligned roughly perpendicular to the enveloping filament (see Figure \ref{fig.m12i_density} for the case of \textbf{m12i}). These observations are consistent with the findings of the DM-only Aquarius simulations as detailed by \citep{vera2011}, where they identify a trend of halos with short axes (at the virial radius) aligning perpendicular to the filament direction.

\begin{figure*}
    \centering
    \textbf{m12f}\par\medskip
    \includegraphics[width=.9\textwidth]{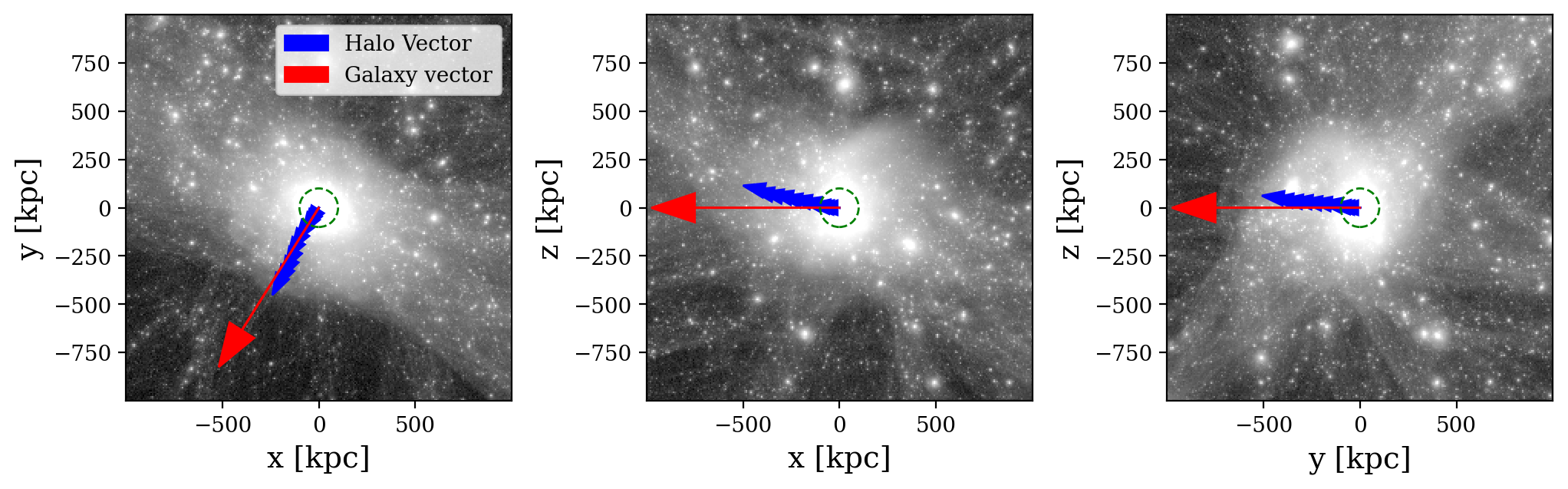} \\
    \textbf{m12i}\par\medskip
    \includegraphics[width=.9\textwidth]{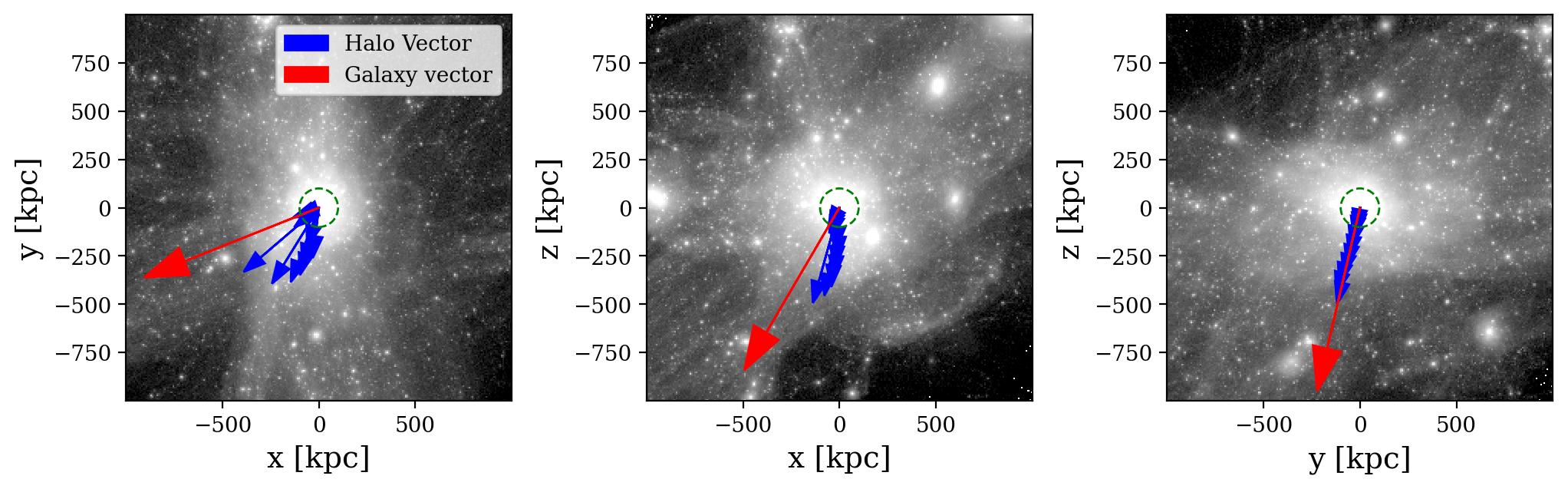} \\
    \textbf{m12m}\par\medskip
    \includegraphics[width=.9\textwidth]{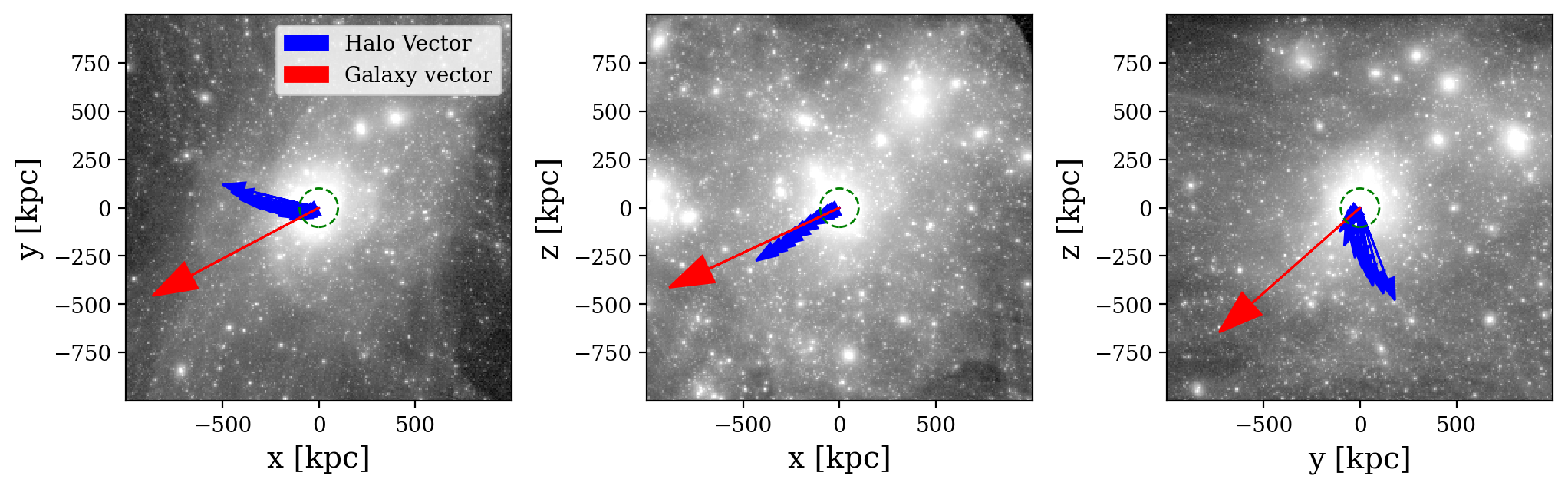} \\
    \textbf{m12w}\par\medskip
    \includegraphics[width=.9\textwidth]{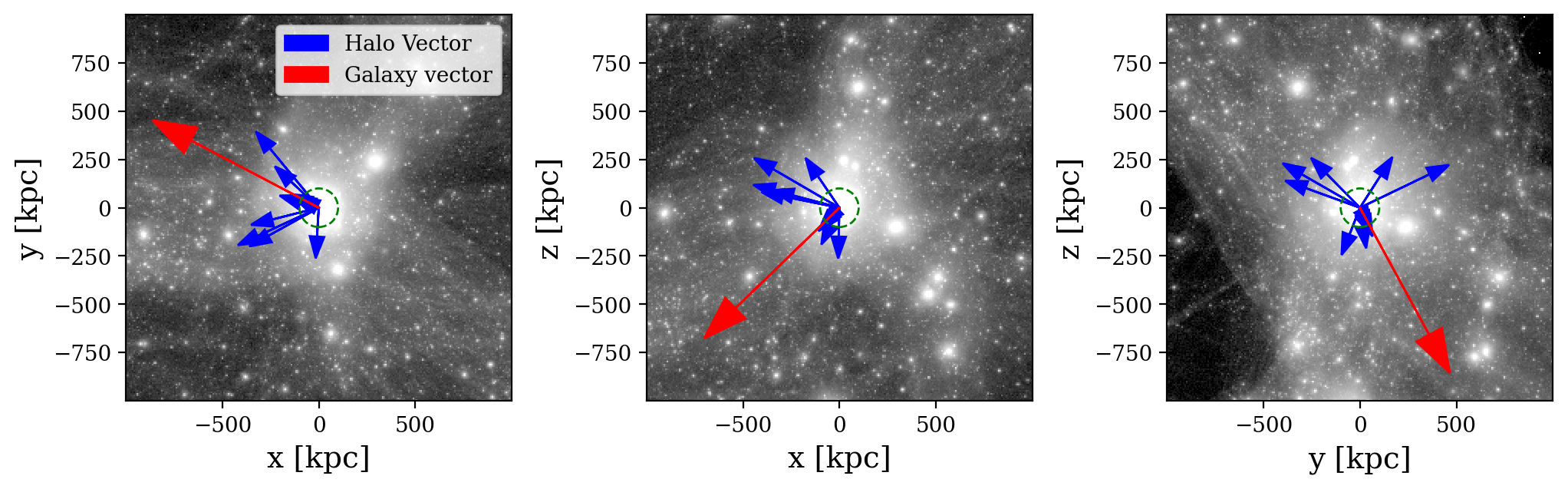}
    \caption{The dark matter densities within each Latte CDM galaxy. The red arrow indicates the direction of the stellar minor axis. The blue arrows indicate the direction of the DM halo minor axis with the length scaled to the radius at which the axes are determined. The dashed green inner circle indicates a radius of 100 kpc.}
    \label{fig:m12_allaxes}
\end{figure*}

This pattern of halo-filament orientation has a critical implication: in-falling substructure mass will preferentially accrete along the major axis of the halo \citep{vera2011}. In the context of \textbf{m12w} we observe significant galaxy-halo decoupling in the outer halo (likely due to the major merger), which indicates that the galaxy may become independently oriented to the filament. In Figure \ref{fig:m12_allaxes}, we observe that \textbf{m12w} is located at the nexus of three filaments that are accreting massive subhalos near the central galaxy. All the other Latte galaxies appear to be roughly perpendicular to the main accreting filaments they are embedded within, however, \textbf{m12w} experiences subhalo accretion from multiple directions causing the DM halo axes to twist significantly as numerous in-falling substructures enter the selection aperture between the twist radius ($R \sim 100$ kpc) out to the virial radius. Future analysis of the environment of the filamentary substructure and its relation to galaxy/outer-halo orientation in Latte galaxies may be critical in understanding the mechanisms behind this observed decoupling.

\subsection{Formation of the Thin Disk and the Stability of the Galaxy Direction}

Figure \ref{fig.m12cdm_vsr} presents the rotational support ratio $\langle v_\phi \rangle / \sigma_{v,\phi}$ of cold HI gas in each galaxy and the approximate time at which the thin disk appears ($t_{\rm disk}$). We find 3 out of 4 galaxies in our CDM+Baryons sample form their thin disks after the first pericentric passage of the LMC analog. The cold gas component that eventually forms into the thin disk is deposited on the same orbital vector as the massive merging satellite \citep{santistevan21}, and we expect that the thin disk imprints the direction of the merger and, by loose correlation, the filament direction at time of the merger.

From our results in Figures \ref{fig.m12_orientations_z} and \ref{fig.m12_orientations_z_at_z0}, it appears that the direction of the stellar component is stabilized due to the merger, and when the thin disk forms it is oriented along the direction of the merger’s angular momentum while simultaneously the inner DM halo strongly aligning to this stellar axis configuration.

We believe that deviations or twists from the inner halo (which should be set by the stellar disk) to the outer halo are attributed to the evolution of the filament direction. For example, \textbf{m12w} is presently accreting massive substructure along three different filamentary directions---most likely causing the DM symmetry axes to distort as those substructures accrete into the moment of inertia aperture.

\section{Conclusions}
\label{s.conclusions}

In this work, we investigated the orientations of the DM halo symmetry axes as a function of radii and time for DM, SIDM, and DMO halos. Furthermore, we explored how mid-scale to major mergers ($1/10 - 1/3\, {\rm M}_{\rm{MW}}$) can influence these symmetry axes in terms of orientation relative to the stellar disk and halo alignment to the LMC. The main conclusions we can draw from this analysis are:

\begin{itemize}
    \item At $z=0$, the orientations of the dark matter halo are well-aligned to the stellar disk at central to intermediate radii (0.2-0.4 $R_{\rm 200m}$) and diverge the greater radii ($>0.4 \, R_{\rm 200m}$) (Figure \ref{fig.m12_triax_orientation_z0}). These results are consistent with the fact that there is no a priori reason that the DM halo should be aligned with the disk axes at all radii. Galactic potential models that assume DM-stellar alignment may be insufficient in describing the true global potential in the outer halo.
    
    \item The symmetry axes of MW-mass DM halos diverge from those within the central galaxy (exceed 20 degrees of obliquity) well within the virial radius ($\sim 0.2-0.4 \, R_{\rm 200m}$). Using asteroseismologically determined distances (using M-giants) combined with instruments such as Roman can give us the necessary observations required to detect streams far enough in the MW to detect divergences in the halo orientation.
    
    \item In the presence of an in-falling LMC-analog, we find differential major axis-LMC alignment behavior depending on the mass of the analog satellite. In Figure \ref{fig.m12_lmc_all}, galaxies undergoing a major LMC merger (\textbf{m12f}, \textbf{m12w}) have their DM major axes align to the satellite as in-fall progresses. However, galaxies undergoing a minor LMC merger do not have an appreciable effect on the direction of the major axes. 
    
    \item When viewing the alignment as a function of radius, we see that major-merging galaxies show a distinct radius-alignment relationship (Figure \ref{fig.align_r}). This apparent trend is not replicated in minor-merging galaxies. We speculate that the lower-mass LMC satellites do not generate a significant amount of torque on the halo to strongly shift the dark matter density, leaving the axes roughly stationary (see Figure \ref{fig.m12_lmc_all}). 
    
    \item The outer halo more closely aligns to the LMC-analog position over time than the inner halo, showing the presence of differential response of the halo relative to the LMC across galactic radii. 
    
    \item In simulations initialized with SIDM (DMO), the orientations to the stellar disk (pseudo-stellar disk) are more aligned compared to the orientations of CDM halos with baryons. We predict this effect to be stronger in dissipative SIDM models.
    
    \item We find evidence that the orientation of the thin disk is set by the orbital direction of the merger, and that the inner DM halo simultaneously couples to the stellar axes with the same configuration. Deviations to these axes in the outer halo are believed to be caused by changes in the direction of substructure accretion due to the evolution of filamentary structure. 
    
\end{itemize}

 An observational test of the halo orientation can be useful in explaining phenomena such as the figure rotation of the DM halo \citep{valluri21}. A conceivable observational method for determining the halo orientation may come from fitting the Milky Way DM distribution function using data from RR Lyrae stars \citep{hattori20}. This method has been used to fit the halo shape within the disk using observational data from Gaia, however, assumes a perfectly aligned halo \citep{hattori20}. Extending this method to tracer stars beyond the baryonic potential of the disk could be useful in constraining the true Milky Way orientation.

Next-generation observations of the Milky Way will reach the orientation transition radius by enabling the detection of these faint tracers in the outer halo. In Figure \ref{fig.m12_triax_orientation_z0}, we indicate the different distance ranges for different instruments as well. The Gaia satellite, when observing giant branch stars, is limited to observations at ~60 kpc \citep{whitepaper}. However, using ground-based asteroseismology measurements of M-giants, as detailed in \cite{astero}, we may be able to measure stellar (and inherently stream) features up to 200 kpc. We see that in \textbf{m12w} (a case of extreme obliquity), we may be able to resolve this orientation signal through observation. Additionally, the next-generation LSST and its instruments can provide us with up to 400 kpc (beyond the virial radius) of distance and would be more than capable of resolving the stream features necessary to detect divergences in the halo orientation \citep{whitepaper}.

Critical conclusions from future observations and simulations can help resolve theories on Modified Newtonian Dynamics (MOND), where an oblique halo relative to the disk challenges a requirement of MOND that the net potential of the halo and the disk should be well-aligned \citep{debattista}. Thus, continuing our investigations of simulated DM halos can provide us with powerful predictions of outer halo dynamics in advance of future observational studies.

\section*{Acknowledgements}
JB acknowledges support from the Research Experience for Undergraduates program at the Institute for Astronomy, the University of Hawaii-Mānoa funded through NSF grant \#2050710. JB thanks the Institute for Astronomy for their hospitality during the development of this project. 

JB, RES, SC, and DH gratefully acknowledge support from NSF grant AST-2007232.

RES additionally acknowledges support from NASA grant 19-ATP19-0068, from the Research Corporation through the Scialog Fellows program on Time Domain Astronomy, and from HST-AR-15809 from the Space Telescope Science Institute (STScI), which is operated by AURA, Inc., under NASA contract NAS5-26555.

MBK acknowledges support from NSF CAREER award AST-1752913, NSF grants AST-1910346 and AST-2108962, NASA grant 80NSSC22K0827, and HST-AR-15809, HST-GO-15658, HST-GO-15901, HST-GO-15902, HST-AR-16159, and HST-GO-16226 from STScI.

CAFG was supported by NSF through grants AST-1715216, AST-2108230,  and CAREER award AST-1652522; by NASA through grants 17-ATP17-0067 and 21-ATP21-0036; by STScI through grants HST-AR-16124.001-A and HST-GO-16730.016-A; by CXO through grant TM2-23005X; and by the Research Corporation for Science Advancement through a Cottrell Scholar Award.

AW received support from: NSF via CAREER award AST-2045928 and grant AST-2107772; NASA ATP grant 80NSSC20K0513; HST grants AR-15809, GO-15902, GO-16273 from STScI.

Support for PFH was provided by NSF Research Grants 1911233, 20009234, 2108318, NSF CAREER grant 1455342, NASA grants 80NSSC18K0562, HST-AR-15800. Numerical calculations were run on the Caltech compute cluster ``Wheeler,'' allocations AST21010 and AST20016 supported by the NSF and TACC, and NASA HEC SMD-16-7592.


This project was made possible by the computing cluster resources provided by the Flatiron Institute Center for Computational Astrophysics.

\section*{Data Availability}
The FIRE-2 simulations are publicly available \citep{Wetzel2022} at \url{http://flathub.flatironinstitute.org/fire}.
Additional FIRE simulation data is available at \url{https://fire.northwestern.edu/data}.
A public version of the \textsc{Gizmo} code is available at \url{http://www.tapir.caltech.edu/~phopkins/Site/GIZMO.html}.

Latte snapshots for \textbf{m12i}, \textbf{m12f}, and \textbf{m12m} are also available at \url{ananke.hub.yt} and \url{binder.flatironinstitute.org/~rsanderson/ananke}.

The code and the associated data files generated as part of this analysis is available on Github at \url{https://github.com/jaybaptista/latte_orientations}.

\section{Appendix}


\subsection{LMC Infall Tracking}

\begin{figure*}[htbp]
    \centering
     \caption{\textbf{Upper panel:} Position of the LMC satellite (circle) relative to the galaxy (origin) and the direction of the galaxy's major axis (cross-hair, exaggerated scale). The color indicates the snapshot the points were taken from (lightest color corresponds to $z \sim 0$). \textbf{Lower panel:} Angle between the major axis to the LMC position.}
    \includegraphics[width=.9\textwidth]{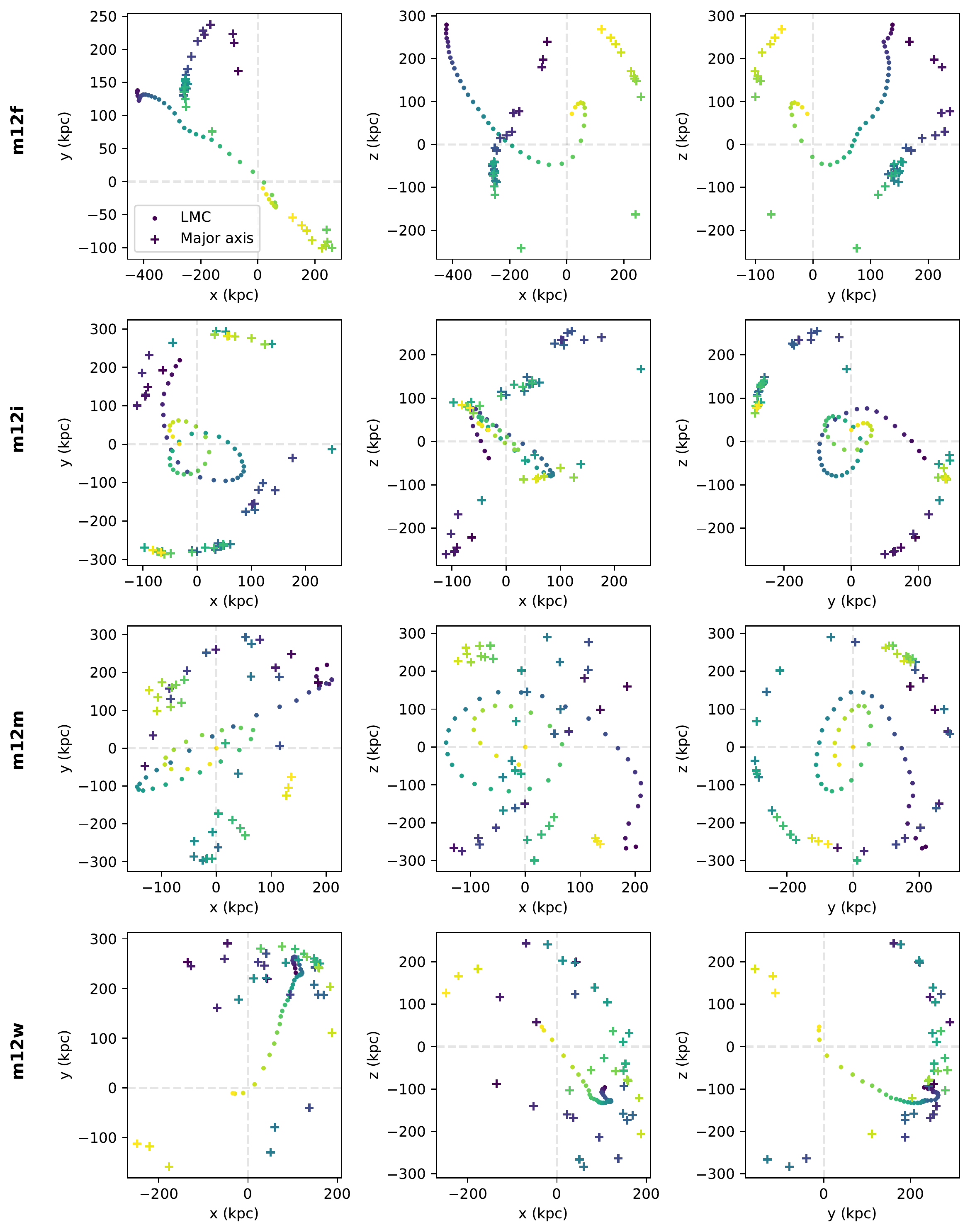}
    \label{fig.m12_lmc_all}
\end{figure*}

Figure \ref{fig.m12_lmc_all} shows the LMC positions and the direction of the major axis vector of each galaxy. 

\subsection{Halo DM Density Visualization}

Figure \ref{fig.m12_lmc_all} shows the LMC positions and the direction of the major axis vector of each galaxy. 

\bibliography{paper}{}
\bibliographystyle{aasjournal}

\end{document}